\documentclass[usenatbib]{mn2e}
\usepackage{epsfig}
\usepackage{amsmath}
\usepackage{subfig, amssymb, color}

\def\msun{{\rm M_{\odot}}}

\def\lsun{{\rm L_{\odot}}}

\newcommand{\be}{\begin{equation}}
\newcommand{\ee}{\end{equation}}

\def\ltsima{$\; \buildrel < \over \sim \;$}
\def\simlt{\lower.5ex\hbox{\ltsima}}
\def\gtsima{$\; \buildrel > \over \sim \;$}
\def\simgt{\lower.5ex\hbox{\gtsima}}
\def\sgra{Sgr~A$^*$}

\def\del#1{{}}

\title{AGN outflows trigger starbursts in gas-rich galaxies}

\author[Kastytis Zubovas, Sergei Nayakshin, Andrew King, Mark
  Wilkinson]{\parbox{18cm}{Kastytis Zubovas$^{1,2}$, Sergei
    Nayakshin$^1$, Andrew King$^1$, Mark Wilkinson$^1$}\\
$^{1}$ Department of Physics \& Astronomy, University of Leicester, Leicester, LE1 7RH, UK\\
$^{2}$ Center for Physical Sciences and Technology, Savanori\c{u} 231, Vilnius LT-02300, Lithuania\\
{E-mail:~} {\rm kastytis.zubovas@ftmc.lt}}

\begin{document}

\maketitle

\begin{abstract}

Recent well resolved numerical simulations of AGN feedback have shown that its
effects on the host galaxy may be not only negative but also positive. In the
late gas poor phase, AGN feedback blows the gas away and terminates star
formation. However, in the gas-rich phase(s), AGN outflows trigger star
formation by over-compressing cold dense gas and thus provide positive
feedback on their hosts. In this paper we study this AGN-triggered starburst
effect. We show that star formation rate in the burst increases until the star
formation feedback counteracts locally the AGN outflow compression. Globally,
this predicts a strong nearly linear statistical correlation between the AGN
and starburst bolometric luminosities in disc galaxies, $L_* \propto L_{\rm
  AGN}^{5/6}$. The correlation is statistical only because AGN activity
  may fluctuate on short time scales (as short as tens of years), and because
  AGN may turn off but its effects on the host may continue to last until the
  AGN-driven outflow leaves the host, which may be up to 10 times longer than
  the duration of the AGN activity. The coefficient in front of this relation
depends on the clumpiness and morphology of the cold gas in the galaxy. A
  ``maximum starburst'' takes place in a azimuthally uniform gas disc, for
  which we derive an upper limit of $L_* \sim 50$ times larger than $L_{\rm
  AGN}$ for typical quasars. For more clumpy and/or compact cold gas
distributions, the starburst luminosity decreases. We also suggest that
  similar AGN-triggerred starbursts are possible in hosts of all geometries,
  including during galaxy mergers, provided the AGN is activated. Finally, we
  note that due to the short duration of the AGN activity phase the
  accelerating influence of AGN on starbursts may be much more common than
  observations of simultaneous AGN and starbursts would suggest.
\end{abstract}

\begin{keywords}
{accretion, accretion discs --- quasars:general --- black hole physics ---
  galaxies:evolution --- stars:formation}
\end{keywords}

\section{Introduction}

Supermassive black holes (SMBHs) are known to reside in the centres of most
large galaxies. The masses of these SMBHs correlate with a range of properties
of their host spheroids, including luminosity \citep[e.g.][]{Magorrian1998AJ}
and mass \citep[e.g.][]{Marconi2003ApJ, Haering2004ApJ}; and the gravitational
binding energy of the bulge \citep[e.g.][]{Feoli2005IJMPD, Aller2007ApJ}.
Furthermore, the correlation between SMBH mass ($M_{\rm bh}$) and the velocity
dispersion ($\sigma$) of its host spheroid, referred to below as a ``$M_{\rm
  bh}$--$\sigma$ relation'', has been studied by many authors. A power-law fit
to the data, $M_{\rm bh} \propto \sigma^p$, yielded values of $p$ in the range
of $p\sim 4-5$ \citep[cf.][]{Ferrarese2000ApJ, Gebhardt2000ApJa,
  Tremaine2002ApJ, Gultekin2009ApJ}.

The most natural explanation for the observed correlations quoted
above is that it is an imprint of the self-regulated growth of the
SMBH. Winds, other types of outflows and radiation
\citep[e.g.,][]{Silk1998A&A,Fabian1999MNRAS,Ciotti2001ApJ} are
expected to drive the gas out of the host's potential, limiting the
SMBH mass and establishing the observed SMBH-galaxy correlations
\citep{King2003ApJ,King2005ApJ}.

Recently, fast ($v\simgt 1000$ km~s$^{-1}$) molecular outflows
resolved on kpc scales were reported by a number of authors
\citep[e.g.,][]{Feruglio2010A&A,Rupke2011ApJ,Sturm2011ApJ}. These were
interpreted by \cite{Zubovas2012ApJ} as being driven by the quasar
outflow in the energy-driven regime where the shocked primary outflow
does not cool rapidly enough \citep{King2005ApJ, King2011MNRAS,
    Faucher2012MNRASb}. In this scenario the SMBH inflates a huge hot
bubble that sweeps up galactic gas into a shell that is driven out at
velocities a few times that of the host's escape velocity, e.g.,
comparable to what is observed.

\cite{Nayakshin2012MNRASb} numerically simulated the dynamics of the SMBH wind
in three dimensions, and found that when the ambient shocked gas cools
rapidly, the shocked gas is compressed into thin cold dense shells, filaments
and clumps. Driving these high density features out is more difficult
than analytical models predict. Importantly, quasars have another way of
affecting the host in this regime -- by triggering a massive star formation
burst in the cold gas by over-pressurising it. Under these conditions SMBHs
actually accelerate star formation in the host, having a positive rather than
a negative effect on their host galaxies. Note that this does not take away
the quasar's negative feedback role traditionally appreciated in the
literature: \cite{Nayakshin2012MNRASb} find that at later, gas-poor epochs,
when the ambient gas does not cool rapidly, the quasar shock is very effective
in driving the gas outward and thus curtailing further star formation in the
host galaxy.

Here we attempt to quantify the properties of the starburst fuelled by the
pressure of the quasar-driven bubble. Clearly the result of such a calculation
should depend sensitively on the distribution (morphology, clumpiness, total
mass, etc.) of the gas in the host galaxy. To demonstrate our main points, we
make the simplest assumption that the gas resides in a large galactic disc
with properties taken from the models of \citet{Mo1998MNRAS}. We also
assume that the gas in the disc is smoothly distributed. This unrealistic
assumption produces a situation where gas can turn into stars everywhere
  in the disc as soon as it is compressed, provided the cooling time is
  short. Unless the disc accretes material from the galactic halo, this
  situation yields an upper limit to the star formation rate and starburst
  luminosity within the framework of our model. We therefore call such a
  configuration a ``maximum starburst''. A situation of significant accretion
  rate is highly unlikely, because the halo gas is being expelled by the AGN
  outflow and does not cool efficiently. A more realistic system, with cold
gas concentrated in clumps and/or spiral arms, results in a smaller surface
area of the disc that responds to an AGN outflow by increasing its star
formation rate. We consider more general gas geometries in the Discussion
section.

The paper is structured as follows. In Section \ref{sec:pressure} we briefly
review the dynamics of large-scale energy-driven AGN outflows and derive the
pressure inside the outflowing shell. Section \ref{sec:lum} presents the
expected correlation between starburst and AGN luminosity. We then use a
numerical model, which we describe in Section \ref{sec:model}, to calculate
the properties of the triggered star formation in a galactic disc (Section
\ref{sec:sf}). We discuss our results and their implications in Section
\ref{sec:discuss} and summarise them in Section \ref{sec:concl}.

\section{Pressure in an AGN outflow} \label{sec:pressure}

\subsection{Large-scale outflow dynamics}

Pressure inside an accretion disc around an SMBH launches a wind,
which is then accelerated by the Eddington-accreting AGN radiation
pressure to a velocity $v_{\rm w} \sim 0.1c$
\citep{King2003ApJ,King2010MNRASa}. This wind shocks against the
surrounding material and drives an outflow in the AGN host galaxy. The
dynamics of the outflow depend on whether the shocked wind can cool
faster than it expands adiabatically. Efficient cooling can only be
achieved via inverse-Compton scattering on the photons of the AGN
radiation field \citep{Ciotti1997ApJ,King2003ApJ} and happens in the
central parts of the galaxy. Simple estimates give the radius of
transition between cooling and non-cooling regimes as
\begin{equation} \label{eq:rcool}
R_{\rm C} \simeq 520 \; \sigma_{200}\; M_8^{1/2}\; l^{1/2} \left(\frac{f_{\rm
    g}}{f_{\rm c}}\right)^{1/2} \; \mathrm{pc}
\end{equation}
\citep{King2003ApJ,Zubovas2012MNRASb}; here, $\sigma_{200}$ is the
velocity dispersion of the host galaxy in units of $200$~km/s, $M_8$
is the SMBH mass in $10^8 \; \msun$, $l$ is the Eddington luminosity
ratio of the AGN, $f_{\rm g}$ is the ratio of gas density to total
density and $f_{\rm c} = 0.16$ is the cosmological value of $f_{\rm
  g}$. \citet{Faucher2012MNRASb} calculated the cooling radius in
  more detail, accounting for cooling of non-relativistic electrons,
  the nonzero electron-ion energy equilibration timescale and
  non-isothermal background potentials. They find that the cooling
  radius is likely to be even smaller than that given by
  eq. (\ref{eq:rcool}).

On large scales, then, the AGN wind gives up all of its energy to the
outflow. This energy-driven outflow has a kinetic luminosity
\begin{equation}
\dot{E}_{\rm out} = \dot{E}_{\rm w} \simeq 0.05 L_{\rm Edd}
\end{equation}
and drives a forward shock into the ambient medium with a velocity
\begin{equation}\label{eq:vout}
v_{\rm out} = \frac{4}{3} v_{\rm e} \simeq 1230\sigma_{200}^{2/3} l^{1/3}
\left(\frac{f_{\rm g}}{f_{\rm c}}\right)^{1/3}~{\rm km\ s}^{-1}
\end{equation}
\citep{Zubovas2012ApJ}, which clears the spheroidal component of the galaxy on
a timescale much faster than dynamical; $v_{\rm e}$ is the outward velocity of
the contact discontinuity between the shocked wind and the outflow. The mass
outflow rate is
\begin{equation}\label{eq:mout}
\dot{M}_{\rm out} \simeq 3700\sigma_{200}^{8/3} \; \msun\,{\rm yr}^{-1},
\end{equation}
easily able to remove a large fraction of a galaxy's gas on short
timescales. The momentum flow rate of an energy-driven outflow is also larger
than in the wind, by a factor
\begin{equation}\label{eq:massload}
\sqrt{f_{\rm L}} \simeq 21\sigma_{200}^{-2/3},
\end{equation}
where $f_{\rm L} \equiv \dot{M}_{\rm out} / \dot{M}_{\rm w}$ is the mass
loading factor of the outflow. The extra momentum is created by the
acceleration caused by the over-pressurised bubble of shocked wind interior to
the ambient medium.

\subsection{Outflow pressure on dense material} \label{sec:varpres}

\begin{figure}
  \centering
    \includegraphics[width=0.45 \textwidth]{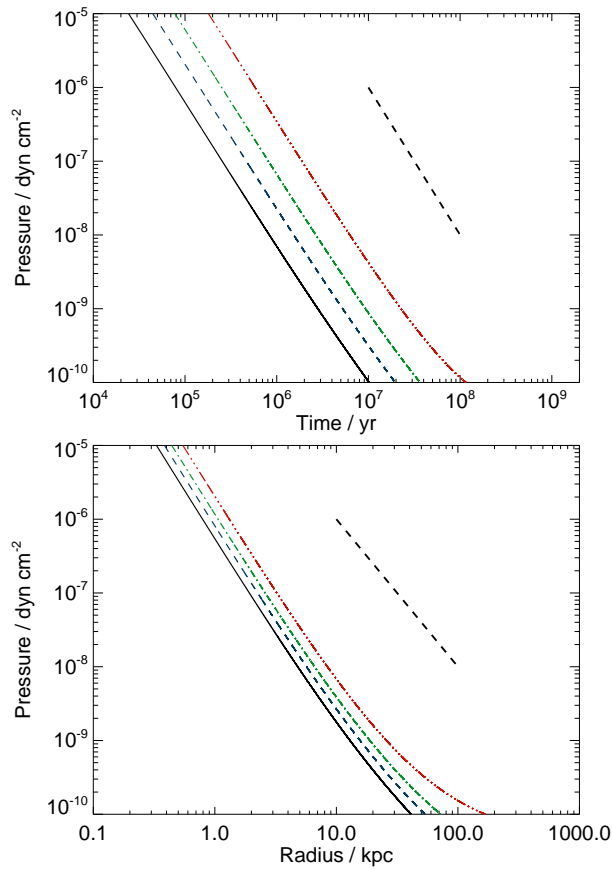}
  \caption{Pressure at the contact discontinuity of the outflow as function of
    time (top panel) and radius (bottom panel) for an NFW background potential
    with $\sigma_{\rm peak} = 200$~km/s, concentration $c = 10$ and scale
    radius $r_{\rm s} = 23$~kpc, for gas fractions of $3 \times 10^{-3}$,
    $10^{-2}$, $3 \times 10^{-2}$ and $0.16$ for the black solid, blue dashed,
    green dot-dashed and red dot-triple-dashed lines, respectively. For
    comparison, typical self-sustained star formation in discs provides
    pressure of $P_* \sim 10^{-11} - 10^{-8}$~dyn cm$^{-2}$. Thick dashed
    lines show power laws with slope $-2$ for easy comparison.}
  \label{fig:pressure}
\end{figure}

The outflow structure is composed of four regions. Moving radially outward from
the centre, these are: free-streaming wind, shocked wind, shocked ambient
medium and undisturbed ambient medium. The pressure in the first and last regions
is not important for our investigation; the free-streaming wind pushes any
material it encounters with at most the ram pressure, which is much lower than
the thermal pressure inside the shocked wind bubble, while the undisturbed
ambient medium is not yet affected by AGN feedback. The pressure in the two
intermediate regions varies between values that can be calculated analytically
for each of the three boundaries: the wind shock, the contact discontinuity
and the outer shock.

The shock front of the wind (the inner shock) is moving backwards with respect
to the contact discontinuity with a velocity $u_{\rm w}$. Using the adiabatic
shock conditions, we have, in the frame of the shock,
\begin{equation}
v_{\rm w} + u_{\rm w} = 4(v_{\rm e} + u_{\rm w}).
\end{equation}
From this, we find
\begin{equation}
u_{\rm w} = \frac{1}{3} v_{\rm w} - \frac{4}{3} v_{\rm e} \simeq \frac{1}{3} v_{\rm w},
\end{equation}
where we used the fact that $v_{\rm w} \gg v_{\rm e}$ in most reasonable
cases. The shock front thus quickly reaches $R_{\rm C}$ from the outside. It
cannot move closer in, because in that region the shock wind cools and loses
its energy very quickly. We thus have a standing shock at $R = R_{\rm C}$. The
pressure at that radius is then
\begin{equation}
P_{\rm C} = \frac{3}{4} \rho_{\rm w} \left(\frac{4}{3} v_{\rm w}\right)^2 =
\frac{\dot{M}_{\rm out} v_{\rm w}}{3 \pi R_{\rm C}^2} \simeq 6 \times 10^{-8}
\sigma_{200}^{-1} \; {\rm dyn} \; {\rm cm}^{-2}.
\end{equation}
We have used equation (\ref{eq:rcool}) to find the numerical value. This
pressure does not vary with time, unless the cooling radius itself varies.

The pressure at the contact discontinuity can be derived from the
energy-driven equation of motion \citep[cf.][]{King2005ApJ}:
\begin{equation} \label{eq:pcd}
P_{\rm CD} = \frac{1}{4 \pi R^2}\left[\frac{{\rm d}}{{\rm
      d}t}\left[M\left(R\right)v_{\rm e}\right] + \frac{G
    M^2\left(R\right)}{f_{\rm g}R^2}\right] = \frac{f_{\rm g} \sigma^2
  \left(v_{\rm e}^2 + \sigma^2\right)}{\pi G R^2}.
\end{equation}
This pressure depends on the radius of the contact discontinuity as $R^{-2}$,
provided that $v_{\rm e}$ is constant with radius. This is the case for an
isothermal potential but is not strictly true in NFW and other more
complicated halos. Numerically, in the case of an isothermal potential,
\begin{equation}\label{eq:pcdnum}
P_{\rm CD} \simeq 4 \times 10^{-7} \sigma_{200}^{10/3} l^{2/3} R_{\rm kpc}^{-2} \;
{\rm dyn} \; {\rm cm}^{-2}.
\end{equation}

Finally, at the outer shock, the pressure is
\begin{equation} \label{eq:pos}
P_{\rm o.s.} = \frac{4}{3} \rho_{\rm amb} v_{\rm e}^2 = \frac{3 f_{\rm g}
  \sigma^2}{8 \pi G R^2} v_{\rm e}^2 \simeq \frac{3}{8} P_{\rm CD}.
\end{equation}
This pressure varies similarly to the pressure at the contact discontinuity.

We show the time variation of pressure at the contact discontinuity (equation
\ref{eq:pcd}) in Figure \ref{fig:pressure}. The precise values of pressure
depend on the assumed background potential; here we consider an NFW profile
with $\sigma_{\rm peak} = 200$~km/s, concentration $c = 10$, scale radius
$r_{\rm s} = 23$~kpc and gas fractions of $3 \times 10^{-3}$, $10^{-2}$, $3
\times 10^{-2}$ and $0.16$ for the black solid, blue dashed, green dot-dashed
and red dot-triple-dashed lines, respectively. The top panel shows pressure
dependence on the time since the start of an AGN activity episode, while the
bottom panel shows the pressure as function of the radius of the contact
discontinuity. For ease of comparison, we draw lines of $P \propto t^{-2}$ and
$P \propto R^{-2}$ (thick dashed lines in the two plots). The pressure
evolution is similar to, but slightly differs from, this simple estimate, due
to the fact that $v_{\rm e}$ is not constant in an NFW profile.

Any structure in a galaxy that is too dense to be blown away by the outflow is
compressed by this external pressure once the outer shock moves past it. For a
galaxy disc, this compression acts perpendicular to the disc plane from both
sides, while in more spherical structures (such as molecular clouds) the
pressure acts from all directions. At any given radius $r$, the evolution of
the external pressure profile can be characterised by three phases. Initially,
before the outflow reaches $r$, the external pressure is small. Once the outer
shock passes $r$, the pressure suddenly increases to the value given by
equation (\ref{eq:pos}) and soon after to the value in equation
(\ref{eq:pcd}). Then, as the outflow bubble expands, the pressure decreases
approximately as
\begin{equation}
P_{\rm ext}\left(t\right) \propto R^{-2} \propto t^{-2},
\end{equation}
as seen in Figure \ref{fig:pressure}, where $t$ is measured from the start of
the AGN activity episode. This phase continues until either the AGN switches
off or until the outflow breaks out past the virial radius of the galaxy and
its velocity starts increasing significantly. In the first case, the pressure
drops more rapidly due to a decrease in outflow velocity $v_{\rm e}$. In the
second case, as can be seen at the bottom of the plots in \ref{fig:pressure},
the pressure decrease slows down as $v_{\rm e}$ starts to rise.

\subsection{Star formation in compressed gas}

Star formation in a galaxy disc takes place when the \citet{Toomre1964ApJ} Q
parameter drops below $\simeq 1$. However, feedback from star formation is
expected to release significant amounts of radiation, mechanical energy and
momentum, heating the disc up. This energy release should increase the
Toomre's parameter, which would then terminate star formation. Therefore, it
is believed that star formation in a galactic disc self-regulates to maintain
the Q-parameter close to unity \citep{Thompson2005ApJ}.  Here we show that in
the presence of a powerful AGN outflow, this self-regulation is significantly
modified, resulting in a much higher star formation rate in the disc.

Qualitatively, we envision the following scenario. A galaxy disc
  is composed of cold gas clouds embedded in warm and hot phases of
  the ISM. The AGN outflow cannot push the disc away radially
  \citep{Nayakshin2012ApJ}, but moves around it through the halo. The
  isotropic pressure in the shocked halo ISM compresses the hot gas in
  the disc. This overpressurized gas exerts approximately isotropic
  pressure upon the cold clouds and drives shockwaves into them,
  leading to an increase in density in the post-shock region
  \citep{Jog1992ApJ}. Gas that was already self-gravitating and on the
  verge of collapsing continues to do so more rapidly, while some gas
  that was unbound becomes self-gravitating due to this density
  increase. Therefore, external pressure can in principle both enhance
  and trigger gravitational collapse in cold gas clouds.

Next, we move on to quantitative estimates. Combining equations
(4) and (6) in \citet{Thompson2005ApJ}, we derive the midplane
pressure of the self-regulated $Q\approx 1$ disc:
\begin{equation}
P_{\rm disc} = \rho_{\rm disc} c_{\rm s}^2 = \frac{1}{2 \sqrt{2}}\frac{f_{\rm
    d}^2 \sigma^4 Q}{\pi G R^2},
\end{equation}
where $f_{\rm d}$ is the ratio of disc gas mass to total dynamical
mass and $c_{\rm s}$ is the sound speed in the disc. Comparing this
with the outflow pressure at the outer shock (eq. \ref{eq:pos}), we
obtain the ratio
\begin{equation}
\frac{P_{\rm o.s.}}{P_{\rm disc}} = \frac{3}{2\sqrt{2} Q} \frac{f_{\rm
    g}}{f_{\rm d}^2} \frac{v_{\rm e}^2}{\sigma^2} \sim 100 \left(\frac{f_{\rm
    g}}{f_{\rm c}}\right)^{5/3} \sigma_{\rm 200}^{-2/3},
\end{equation}
where we assume $f_{\rm d}^2 \simeq f_{\rm g} = f_{\rm c}$. We see, therefore,
that the AGN outflow pressure is one or two orders of magnitude higher than the
internal pressure of the disc, independently of radius, and thus the disc is
significantly compressed.

The response of the galactic disc to this compression can be
  estimated as follows. External pressure on the disc compresses the
  disc and star-forming clouds within it. The disc gas cools rapidly
  \citep[Appendix B]{Thompson2005ApJ} and contracts, increasing gas
  density and the star formation rate. The disc SFR thus increases
  until the turbulent pressure created by stellar feedback counteracts
  not only the disc self-gravity, but the external pressure as
  well. As a result, the SFR increases above the values calculated in
  \citet{Thompson2005ApJ} until it saturates at
\begin{equation} \label{eq:sfrdens}
\dot{\Sigma}_* \simeq \frac{P_{\rm o.s.}}{\epsilon c} \simeq 2.4
\times 10^3 \; \epsilon_{-3} \sigma_{200}^{10/3} l^{2/3} R_{\rm
  kpc}^{-2} \; \msun \; {\rm kpc}^{-2} \; {\rm yr}^{-1},
\end{equation}
where $\epsilon \sim 10^{-3}\epsilon_{-3}$ is the efficiency of
  mass-to-radiation conversion by massive stars
  \citep{Leitherer1992ApJ}. This approach assumes a quasi-steady state
  in which massive stars end their life cycle in supernova explosions
  at the same rate as new massive stars are born.

Clearly, the steady-state situation assumed by equation
  \ref{eq:sfrdens} sets in only after $2-4$~Myr, the lifetime of the
  most massive stars. One may therefore think that at earlier times
  the star formation rate density could actually significantly exceed
  this limit since supernova explosions from massive stars have not
  yet occured. However, winds from massive stars provide radiation
  pressure support against collapse as well. \citet{McLaughlin2006ApJ}
  showed that young stellar clusters produce feedback with a thrust
  $\dot{p}_* \simeq 0.05 L_{\rm Edd}/c$, where $L_{\rm Edd}$ is the
  Eddington luminosity for the cluster mass. Using the definition of
  Eddington luminosity and the fact that $P_* = \dot{p}_* / A$, where
  $A$ is the area, we find that
\begin{equation} \label{eq:sigmamax}
\Sigma_{\rm *,max} \simeq \frac{10 \kappa P_{\rm o.s.}}{\pi G} \simeq
1.4 \times 10^{10} \; \sigma_{200}^{10/3} l^{2/3} R_{\rm kpc}^{-2} \;
\msun \; {\rm kpc}^{-2};
\end{equation}
an extra factor of two comes from the fact that external pressure
  acts on both sides of the disc. This equation shows that the initial
  burst of rapid star formation terminates when the surface mass
  density of the young stellar population builds up above $\sim 1.5
  \times 10^{10} \; \msun$ kpc$^{-2}$. As massive stars die, their
  stellar wind feedback decreases, and thus new stars must be born to
  maintain the pressure balance. This brings us back to the
  steady-state limit given by equation \ref{eq:sfrdens}.

To show that the initial pre-supernova star burst does not exceed
  the steady-state star formation rate significantly, we assume an
  average massive star lifetime of $10$~Myr, and write an approximate
  maximum star formation rate density
\begin{equation} \label{eq:sigmadotmax}
\dot{\Sigma}_{\rm *,max} \simeq \frac{\Sigma_{\rm *,max}}{10 {\rm
    Myr}} \simeq 1.4 \times 10^{3} \; \sigma_{200}^{10/3} l^{2/3}
R_{\rm kpc}^{-2} \; \msun \; {\rm kpc}^{-2} \; {\rm yr}^{-1},
\end{equation}
a value very similar to that obtained in equation
  (\ref{eq:sfrdens}).

 Even averaged over the whole galaxy disc, this SFR density is
 comparable to that in strong starbursts \citep{Chapman2004ApJ} and
 much higher than in quiescent galaxies, such as the Milky Way
 \citep[$\Sigma_{\rm *, MW} \ll 1 \;
   \msun$~kpc$^{-2}$~yr$^{-1}$,][]{Robitaille2010ApJ}.

\section{Relation between AGN and starburst luminosity} \label{sec:lum}

This burst of star formation, proceeding at a rate given by
eq. (\ref{eq:sfrdens}), creates a young stellar population, producing a
radiation flux
\begin{equation} \label{eq:fsb}
F_* = \epsilon_* \dot{\Sigma}_* c^2 \simeq \frac{\epsilon_*}{\epsilon} P_{\rm
  o.s.} c \simeq 3.4 \times 10^{13} \; \sigma_{200}^{10/3} l^{2/3} R_{\rm
  kpc}^{-2} \; \lsun \; {\rm kpc}^{-2};
\end{equation}
here, $\epsilon_*$ is the (time dependent) efficiency of mass-to-radiation
conversion by the whole stellar population. On timescales longer than $\sim
10$~Myr, $\epsilon_* = \epsilon = 10^{-3}$, since we can approximate the
radiation flux to be just that of the young stars, provided that the starburst
is still progressing. On shorter timescales, the precise equality does not
hold, because we must account for the death of massive stars as
well. Therefore a statistical correlation between AGN activity and starburst
luminosity is established in $\sim 10^7$~yr.

We now express the star formation flux in terms of the AGN luminosity. In an
isothermal background potential for an SMBH with $M = M_\sigma = f_{\rm c}
\kappa \sigma^4 / \pi G^2$ \citep{King2010MNRASa} radiating at $l$ times the
Eddington limit, we find
\begin{equation}
\sigma = \left(\frac{GL}{4 f_{\rm c} c l}\right)^{1/4}.
\end{equation}
We substitute this expression into eq. (\ref{eq:fsb}) to find
\begin{equation}\label{eq:flux}
F_* \simeq 8 \times 10^{12} L_{46}^{5/6} l^{-1/6} R_{\rm kpc}^{-2} \; \lsun
\; {\rm kpc}^{-2},
\end{equation}
where $L_{46} \equiv L/10^{46}$~erg/s and we dropped the factor
$\epsilon_*/\epsilon \simeq 1$. The relation between $M_\sigma$ and $\sigma$
is similar for other density profiles, such as NFW, although there $\sigma$
refers to the maximum value of the formal velocity dispersion $\sigma_{\rm
  peak}^2 = G M \left(<r_{\rm peak}\right) / \left(2 r_{\rm peak}\right)$.

In a realistic galaxy, this flux is a local quantity which applies only to
star forming regions. In areas where the cold gas density is initially very
low, the AGN outflow may not be able to trigger star formation at
all. Observationally, this means that if the star forming regions in the
galaxy are not resolved, the observed star formation flux is lower than given
in eq. (\ref{eq:flux}). Therefore, we introduce a factor
\begin{equation}
\xi_{\rm cold} \equiv \frac{\Sigma_{\rm cold}}{\Sigma_{\rm gas}} < 1,
\end{equation}
to account, in a crude fashion, for the clumpiness of the cold gas
distribution. In the ``maximum starburst'' model that we consider for
simplicity, $\xi_{\rm cold} = 1$.

The total luminosity of the starburst is found by integrating the flux and
allowing for cold gas clumpiness:
\begin{equation}
\begin{split}
L_* & \simeq 2 \pi \xi_{\rm cold}\frac{f_{\rm g} \sigma^2 \left(v_{\rm e}^2 +
  \sigma^2\right) c}{\pi G} {\rm ln}\frac{R_{\rm out}}{R_{\rm C}} \\ & \simeq
5 \times 10^{13} L_{46}^{5/6} l^{-1/6} \xi_{\rm cold}{\rm ln}\frac{R_{\rm out}}{0.5
  \sigma_{200}\left(\frac{lf_{\rm g}}{f_{\rm c}}M_8\right)^{1/2}
  \mathrm{kpc}}\; \lsun.
\end{split}
\end{equation}
If we take the outer radius of the starburst to be $\sim 5$~kpc (this radius
is of the same order as scale radii of typical galaxy discs), the luminosity
becomes
\begin{equation} \label{eq:sflum}
L_* \simeq 1.2 \times 10^{14} L_{46}^{5/6} l^{-1/6} \; \lsun = 4.5 \times
10^{47} L_{46}^{5/6} l^{-1/6} \; {\rm erg s}^{-1}
\end{equation}
for typical parameters in the ``maximum starburst'' model. In this model, the
starburst luminosity can surpass that of the AGN by an order of magnitude or
more. The galaxy undergoing such a starburst would appear as a ULIRG.

In a more realistic setting the starburst luminosity may be reduced due to
clumpiness of the cold gas distribution. If $\xi_{\rm cold}$ does not vary by
huge factors between different galaxies, we expect the AGN and starburst
luminosities to be comparable:
\begin{equation}\label{eq:lratio}
\frac{L_*}{L_{\rm AGN}} \approx 4.5 L_{46}^{1/6} \frac{\xi_{\rm cold}}{0.1}
L_{46}^{-1/6} l^{-1/6}.
\end{equation}
We compare this result with observations in Section \ref{sec:lumdiscuss}.

The integrated star formation rate in the galaxy,
\begin{equation}
\dot{M}_* \simeq \xi_{\rm cold} \frac{L_*}{\epsilon c^2} \simeq 10^{3}
\frac{\xi_{\rm cold}}{0.1} L_{46}^{5/6} l^{-1/6} \; \msun \; {\rm yr}^{-1},
\end{equation}
is as large as in the most vigorous starbursts known \citep{Chapman2004ApJ}.

We now investigate the galactic disc response in more detail, using a
numerical model. Structures with different geometry would respond in slightly
different ways, but we do not consider these complications, concentrating on
the properties of the AGN event and how it is reflected in the star formation
within the disc.

\section{Numerical model} \label{sec:model}

The numerical model we use is based on a semi-analytical integrator routine
that follows the dynamics of a spherically symmetric outflow. We combine this
calculation with a 1D grid where we calculate the response of the disc to the
external pressure, including the star formation rate and the luminosity
produced by the starburst and the young stars.

The outflow propagation is calculated using the analytical expressions
from \citet{King2005ApJ} and \citet{Zubovas2012MNRASb}, which are then
integrated numerically for an arbitrary choice of an (analytical)
background potential, black hole mass and luminosity and mean gas
fraction in the bulge and halo of the galaxy. We have previously used
this integrator in \citet{King2011MNRAS} and \citet{Zubovas2012MNRASb}
and have now updated the time-stepping in order to better track outflow
propagation at early times. The numerical results agree perfectly with
the analytical solutions \citep[where these are
  possible][]{King2011MNRAS}.

At each timestep, the pressure of the outflow is calculated using the
analytical expressions from Section \ref{sec:varpres} on a
logarithmically spaced grid with 200 bins spanning a radial range from
$0.4$ to $40$~kpc. We assume that the pressure between the contact
discontinuity and the outer shock varies linearly. The pressure just
outside the cooling radius, however, drops exponentially toward the
value of $P_{\rm CD}$ (eq. \ref{eq:pcd}), with a scale radius $r_{\rm
  s,c} = R_{\rm C} \times v_{\rm out} / v_{\rm w}$. This value is
plausible to within an order of magnitude based on pressure balance,
and the precise numbers are not important to the end result. If the
outflow reaches a maximum radius and begins to collapse, we do not
allow pressure to increase, as this is not physical (a real outflow
would fragment and dissipate rather than collapse
spherically). Rather, we let all pressures drop linearly with time
after the maximum radius is reached, to mimic this dissipation and
radiative cooling of the outflowing material.

The star formation rate density is calculated separately in each bin using
equation (\ref{eq:sfrdens}). Then we calculate the star formation rate
\begin{equation}
\dot{M}_* = 2 \pi r_{\rm bin} \Delta r_{\rm bin} \times \dot{\Sigma}_*,
\end{equation}
where $r_{\rm bin}$ is the radius of the given bin and $\Delta r_{\rm bin} \ll
r_{\rm bin}$ is its width. We then update the total mass formed in that bin
since the start of the simulation and compare this mass with the total gas
mass expected to be available in the pristine disc at that radius; for the
latter, we use the analytical expression for disc gas distribution from
\citet{Mo1998MNRAS}. If $M_* > M_{\rm gas}$, we set $\dot{\Sigma}_* =
\dot{M}_* = 0$ and $M_* = M_{\rm gas}$ in that annulus.

Once we know the correct star formation rate, the mass of newly formed stars
and the stellar surface density, we calculate their emitted flux $F_*$ and
luminosity $L_*$. To do this, we split the stellar population into two
sub-populations: low-mass stars and massive stars. For a Salpeter IMF, the
fraction of mass contained in stars with $m_* > 8 \; \msun$ is $\simeq
10\%$. The low-mass stars are assumed to radiate with a constant luminosity,
and we assign their sub-population a mass-to-light ratio $\Gamma = 1$. Most
massive stars radiate at a good fraction of their Eddington luminosity; we
take that fraction to be $50 \%$, which corresponds to a mass-to-light ratio
$\Gamma \sim 1.7 \times 10^4$, but their mass decreases exponentially on a
timescale of $10$~Myr, simulating their short lifetimes. The flux of both
populations is therefore proportional to their surface densities; the low-mass
star surface density can only grow as new stars are formed, but the massive
stars can both be created and die.

With this setup, we explore a small part of the vast parameter space by
considering five cases of varying outflow and disc properties. In all the
cases, we use the same background NFW potential, described in Section
\ref{sec:pressure}. In the `Base' run, we set the initial SMBH mass equal to
the theoretical $M_\sigma$ value and allow it to grow and radiate at the
Eddington limit for an unlimited time. In the `Low-tq' run, we turn the AGN
off (i.e. set the luminosity and the mass growth of the SMBH to zero) after
$5$~Myr to see what an effect a short burst has. In the `Low-SFR' run, we
reduce $\dot{\Sigma}_*$ by a factor $\xi_{\rm cold} = 0.1$ to allow for gas
clumpiness (see Section \ref{sec:lum}) and other processes that can reduce the
SFR. In the `Low-M0' run, the initial SMBH mass is set to $0.1 \times
M_\sigma$ to find out what happens if a black hole slightly under the formal
mass limit drives a large-scale outflow. Finally, the `Channel' case
represents a situation where the outflow is effectively channelled away from
the plane of the galaxy, only affecting the disc within $4$~kpc of the
centre. We emphasise that we do not intend to cover all possible situations
and configurations with these cases, but merely attempt to show the variations
that can be expected within this model.

\section{Results} \label{sec:sf}

\begin{figure*}
  \centering
    \subfloat{\includegraphics[width=0.32\textwidth]{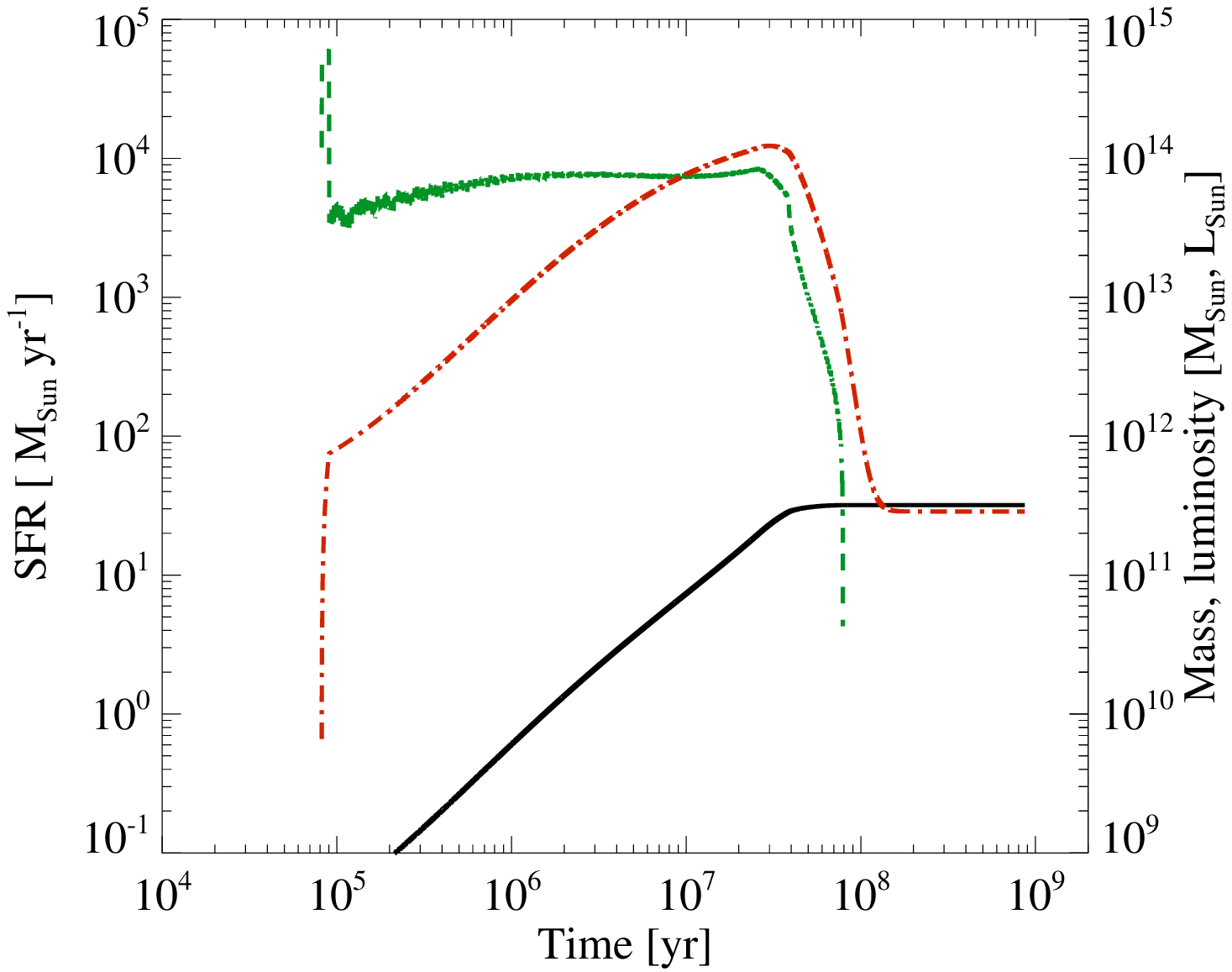}}
    \subfloat{\includegraphics[width=0.32\textwidth]{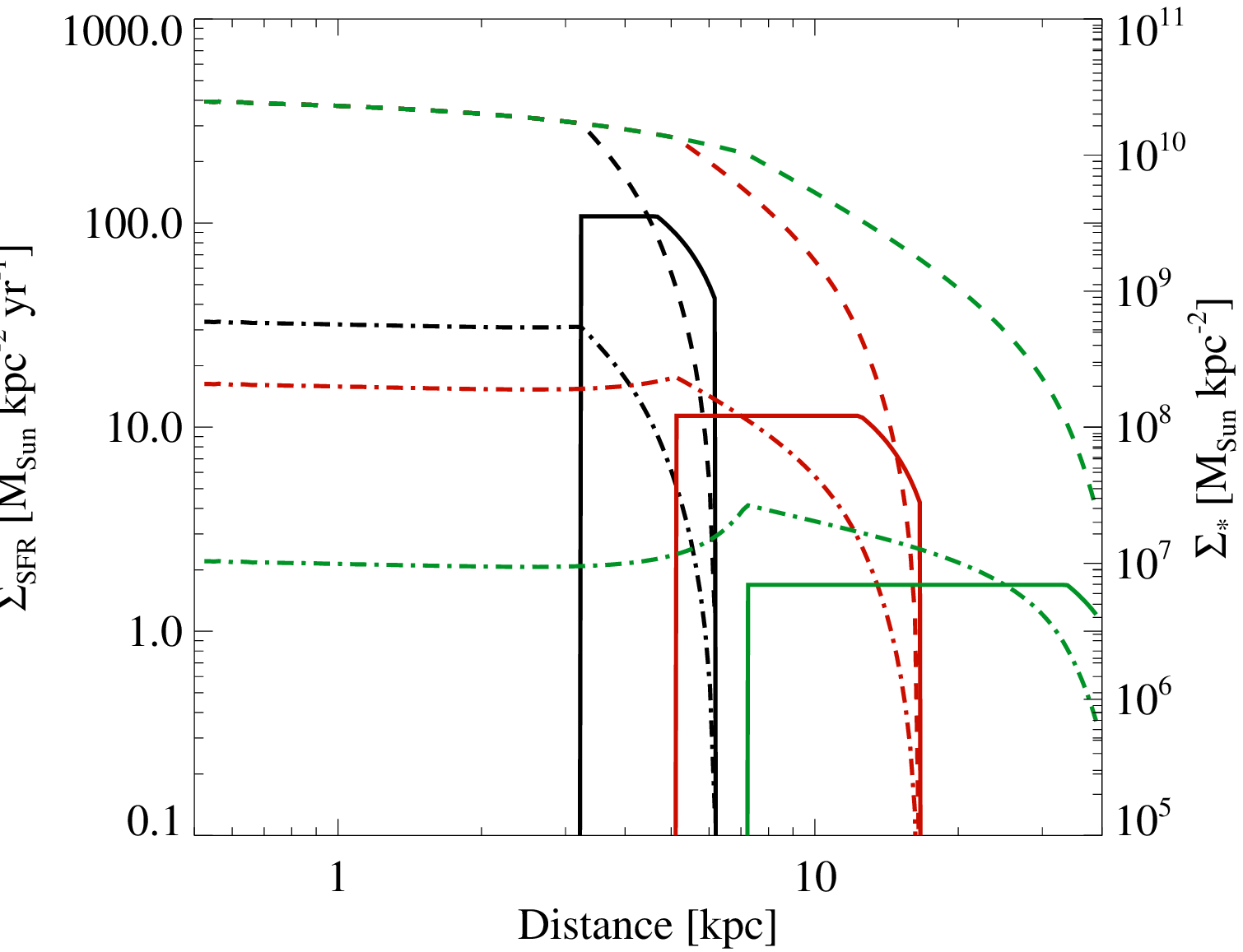}}
    \subfloat{\includegraphics[width=0.32\textwidth]{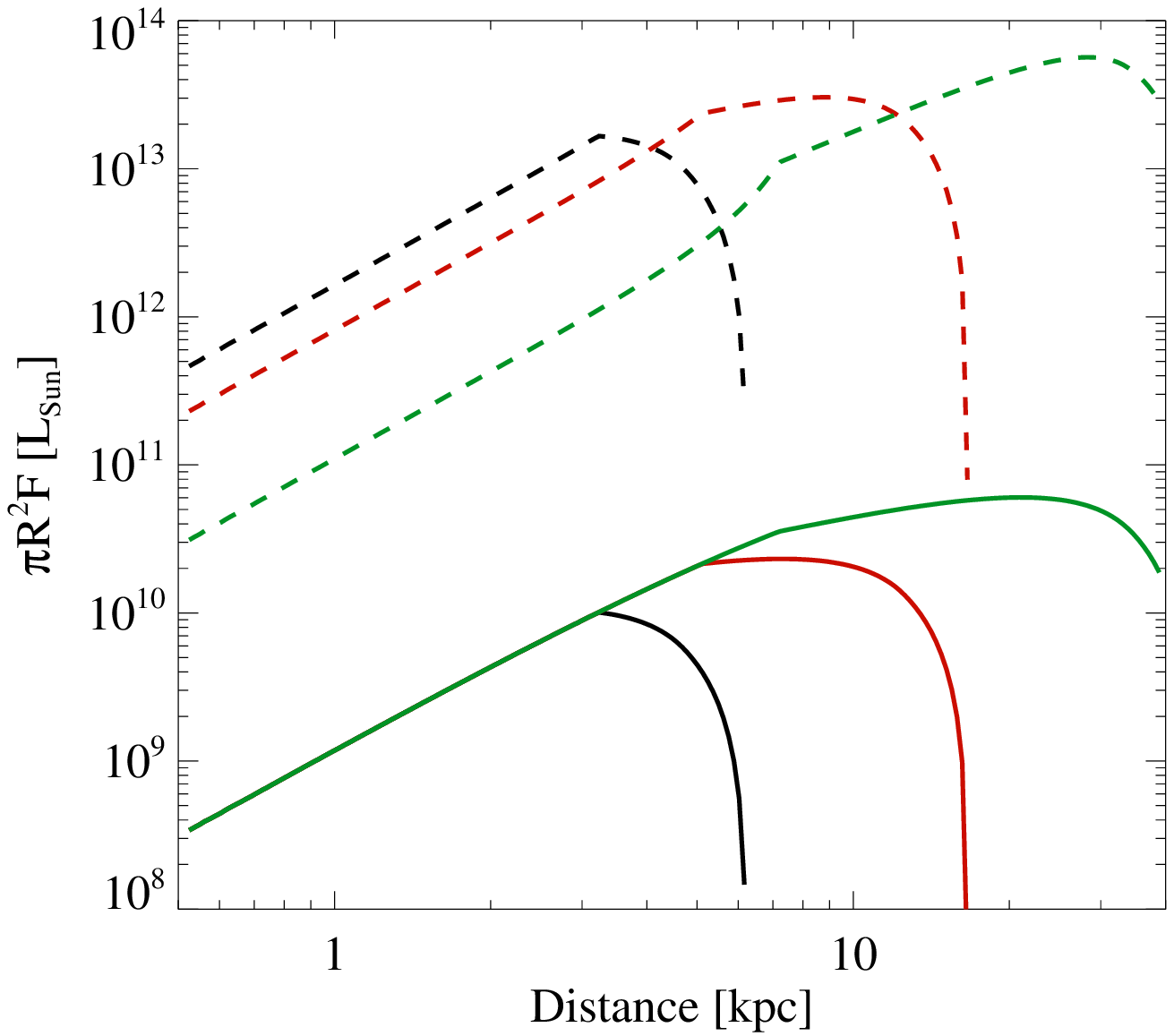}}
  \caption{Properties of the starburst induced by AGN activity in case 1,
    `Base'. The SMBH initially has a mass $3.68 \times 10^8 \; \msun =
    M_\sigma$ and grows and radiates at the Eddington rate. The outflow
    propagates in a static background NFW potential, described in Section
    \ref{sec:pressure}. {\it Left panel}: Time evolution of star formation
    rate (green dashed line, scale on the left), total mass of gas converted
    into stars (black solid line) and luminosity of the newly formed stellar
    population (red dot-dashed line; both scales on the right). The whole gas
    disc, $M_{\rm g} = 3 \times 10^{11} \; \msun$, is converted in
    $50$~Myr. {\it Middle panel}: Radial plots at $3$, $10$ and $30$~Myr
    (black, red and green lines, respectively) of star formation rate density
    (solid, scale on the left), surface density of low-mass stars (dashed) and
    massive stars (dot-dashed; both scales on the right). {\it Right panel}:
    radial plot of $\pi R^2 F_*$ of the low-mass (solid) and massive (dashed
    line) stars; colour coding identical to the middle panel.}
  \label{fig:sfr1}
\end{figure*}

\begin{figure*}
  \centering
    \subfloat{\includegraphics[width=0.32\textwidth]{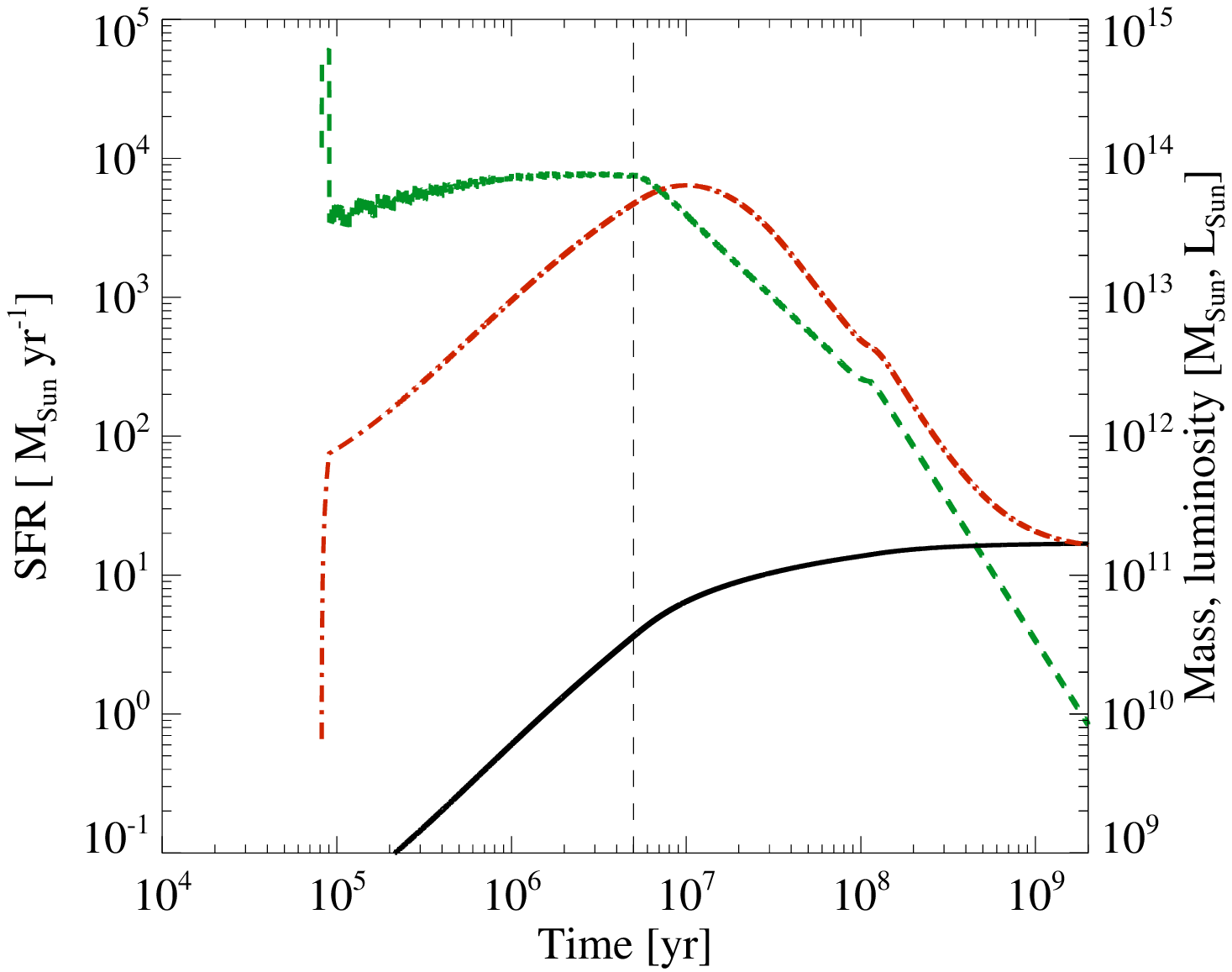}}
    \subfloat{\includegraphics[width=0.32\textwidth]{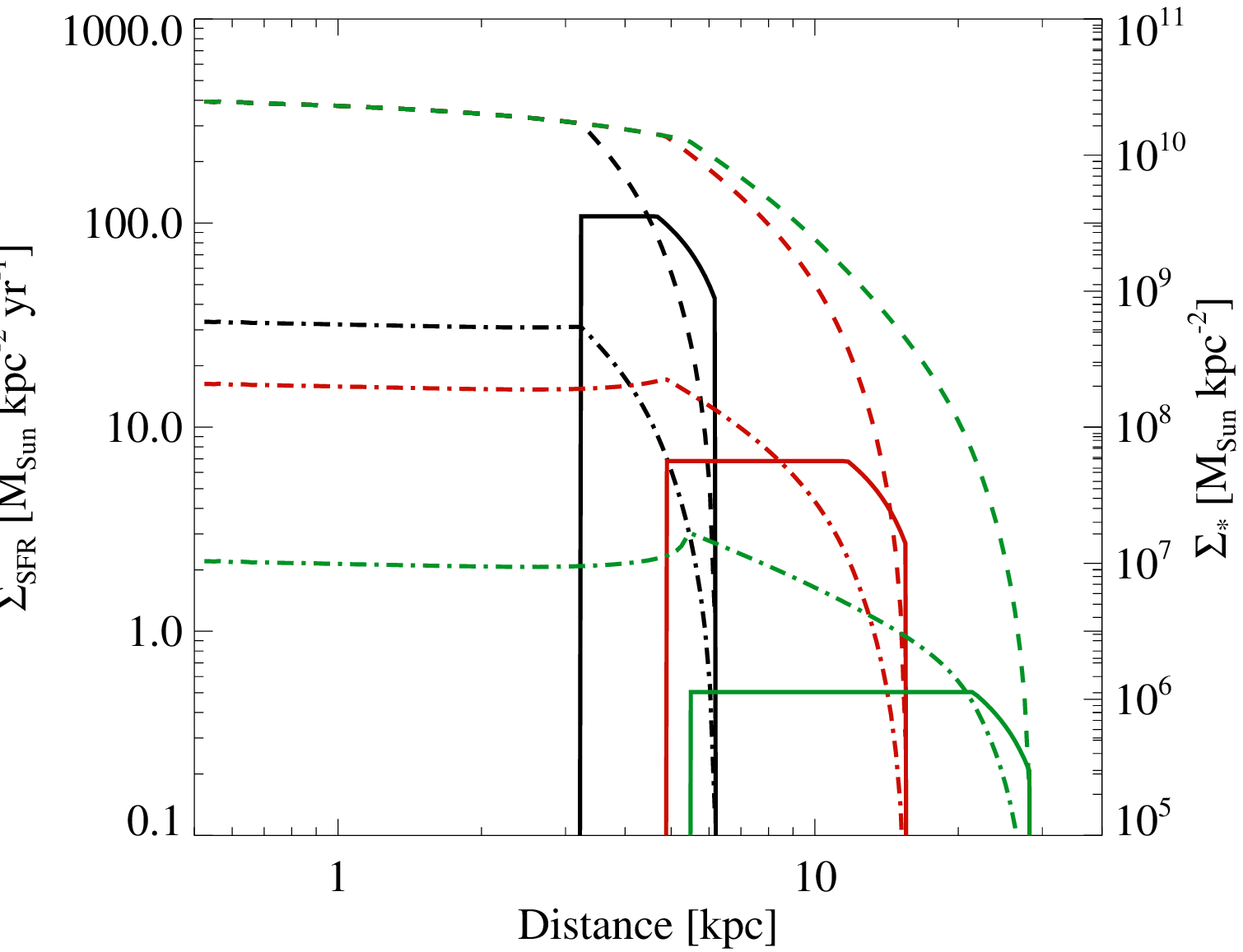}}
    \subfloat{\includegraphics[width=0.32\textwidth]{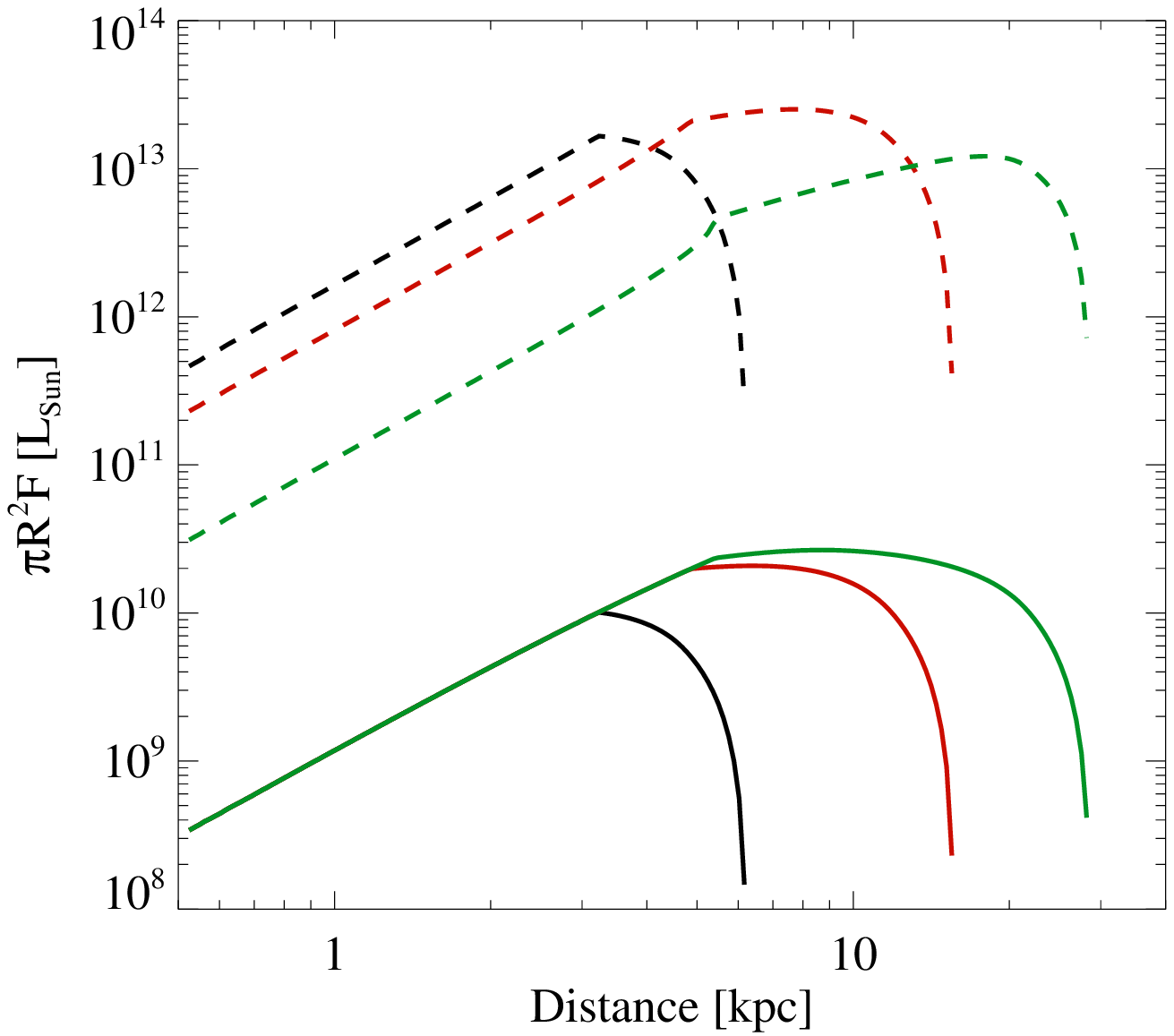}}
  \caption{Case 2, `Low-tq': Same as Figure \ref{fig:sfr1}, but the
    AGN switches off after $t_{\rm q} = 5$~Myr. {\it Left panel}: Time
    evolution of star formation rate (green dashed line, scale on the
    left), total mass of gas converted into stars (black solid line)
    and luminosity of the newly formed stellar population (red
    dot-dashed line; both scales on the right). {\it Middle panel}:
    Radial plots at $3$, $10$ and $30$~Myr (black, red and green
    lines, respectively) of star formation rate density (solid, scale
    on the left), surface density of low-mass stars (dashed) and
    massive stars (dot-dashed; both scales on the right). {\it Right
      panel}: radial plot of $\pi R^2 F_*$ of the low-mass (solid) and
    massive (dashed line) stars; colour coding identical to the middle
    panel.}
  \label{fig:sfr2}
\end{figure*}

\begin{figure*}
  \centering
    \subfloat{\includegraphics[width=0.32\textwidth]{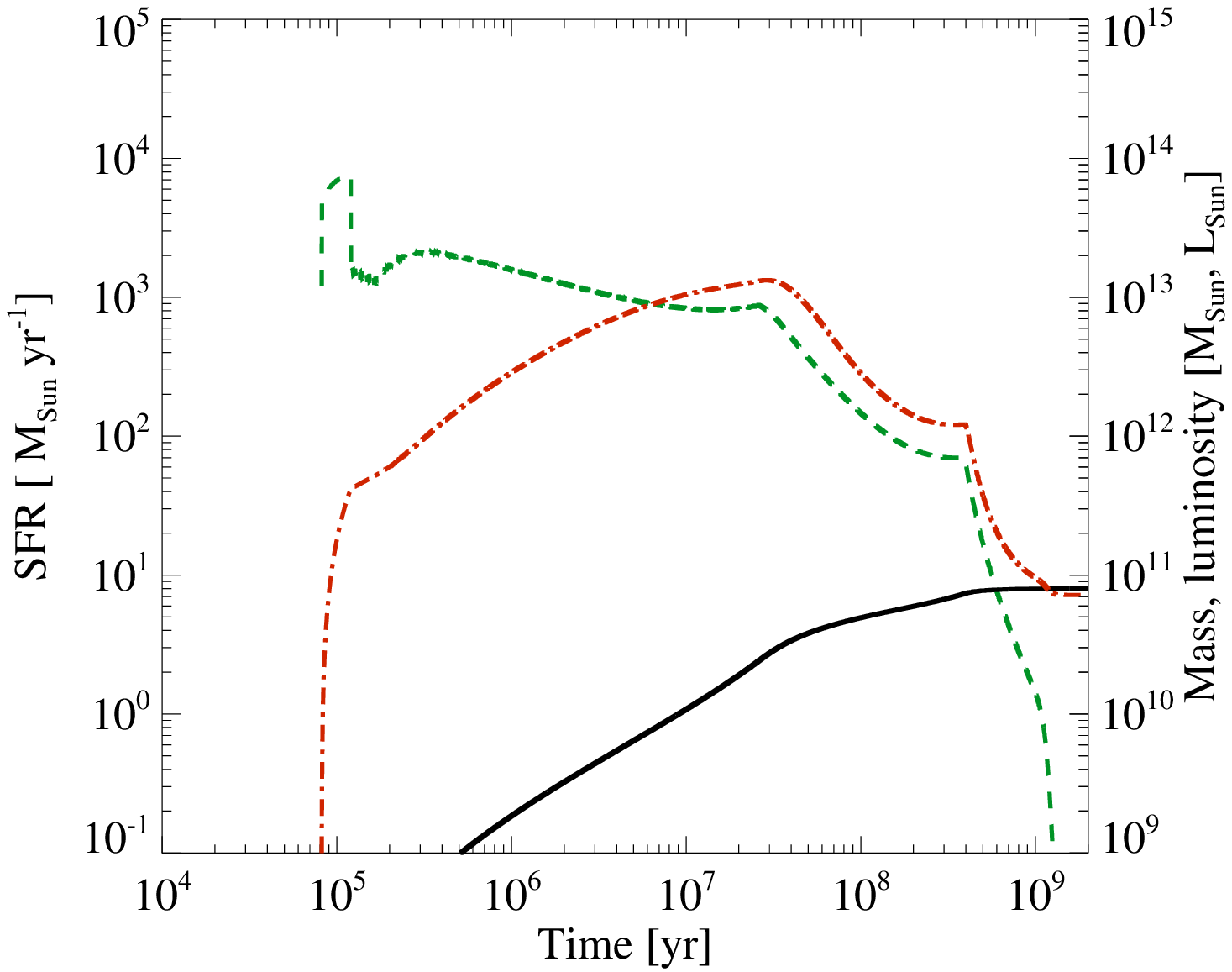}}
    \subfloat{\includegraphics[width=0.32\textwidth]{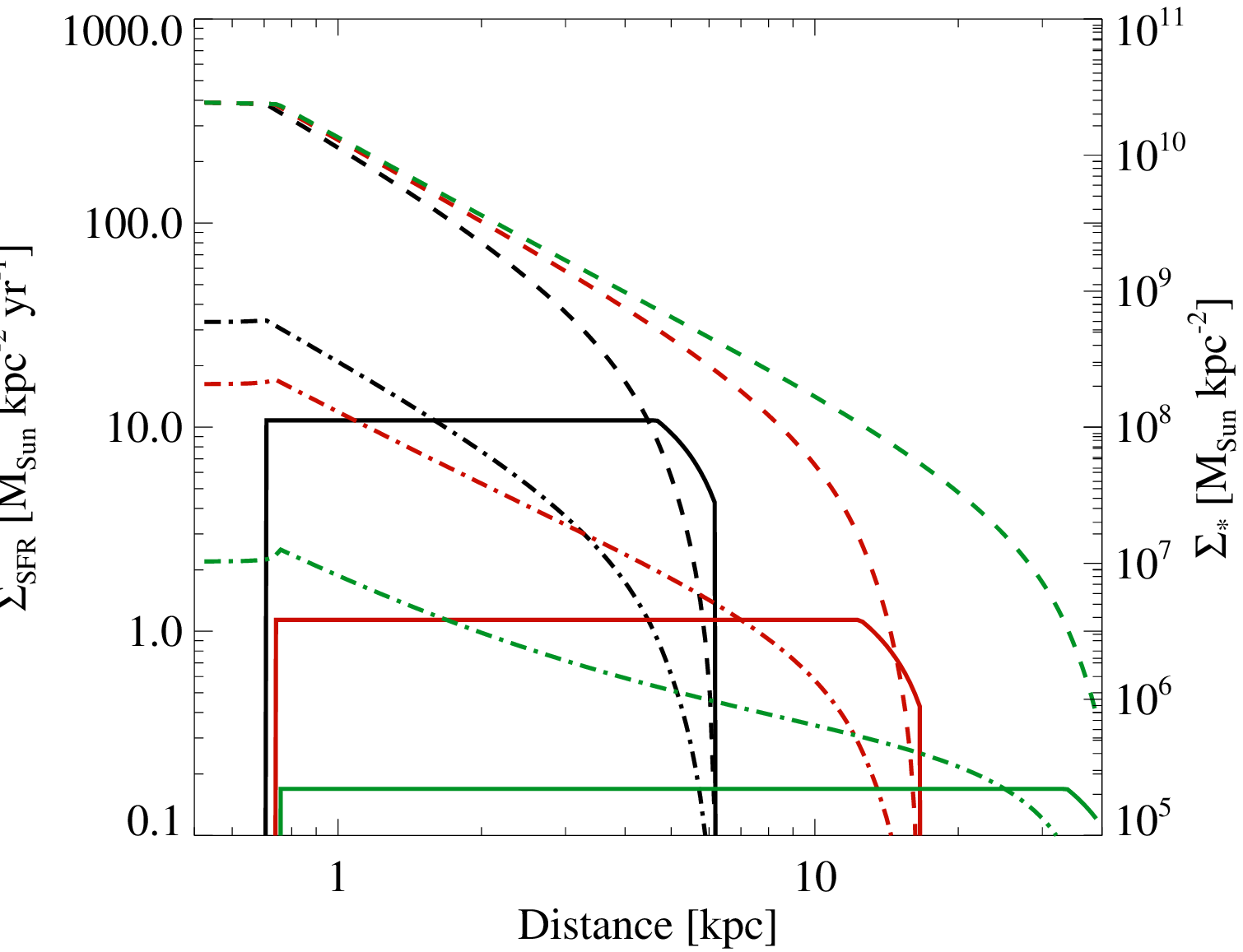}}
    \subfloat{\includegraphics[width=0.32\textwidth]{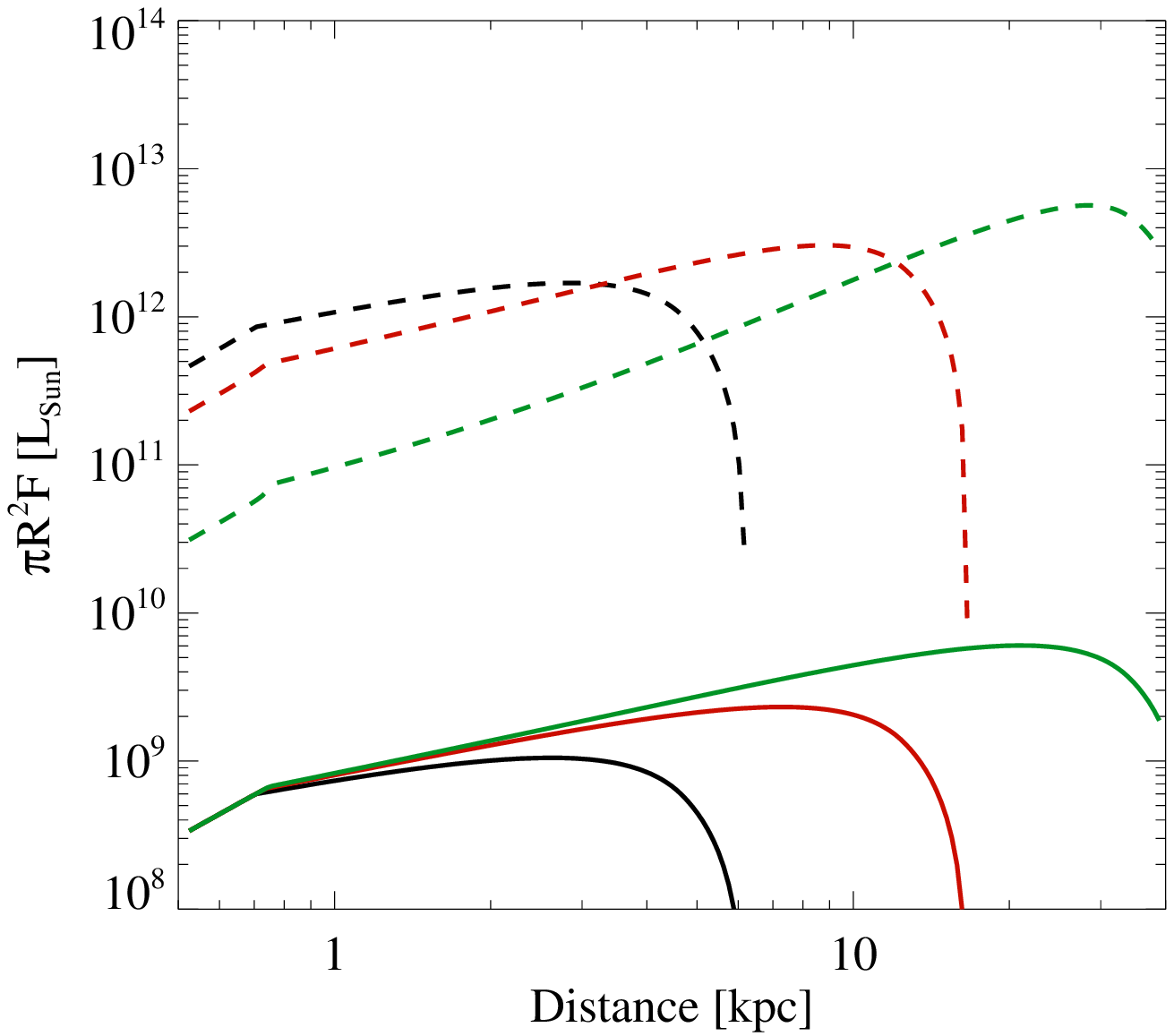}}
  \caption{Case 3, `Low-SFR': Same as Figure \ref{fig:sfr1}, but SFR is
    artificially reduced by a factor 10. {\it Left panel}: Time
    evolution of star formation rate (green dashed line, scale on the
    left), total mass of gas converted into stars (black solid line)
    and luminosity of the newly formed stellar population (red
    dot-dashed line; both scales on the right). {\it Middle panel}:
    Radial plots at $3$, $10$ and $30$~Myr (black, red and green
    lines, respectively) of star formation rate density (solid, scale
    on the left), surface density of low-mass stars (dashed) and
    massive stars (dot-dashed; both scales on the right). {\it Right
      panel}: radial plot of $\pi R^2 F_*$ of the low-mass (solid) and
    massive (dashed line) stars; colour coding identical to the middle
    panel.}
  \label{fig:sfr3}
\end{figure*}

\begin{figure*}
  \centering
    \subfloat{\includegraphics[width=0.32\textwidth]{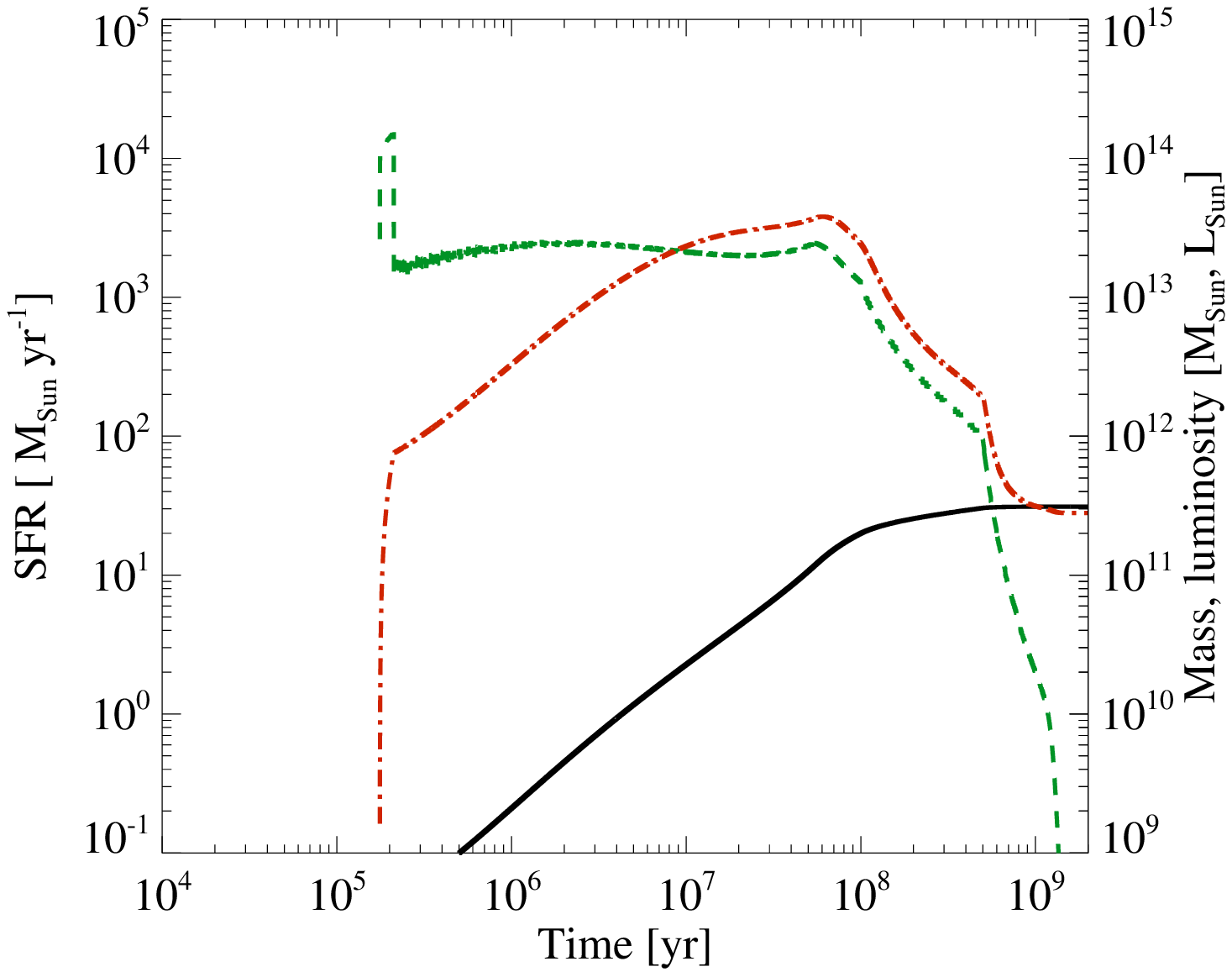}}
    \subfloat{\includegraphics[width=0.32\textwidth]{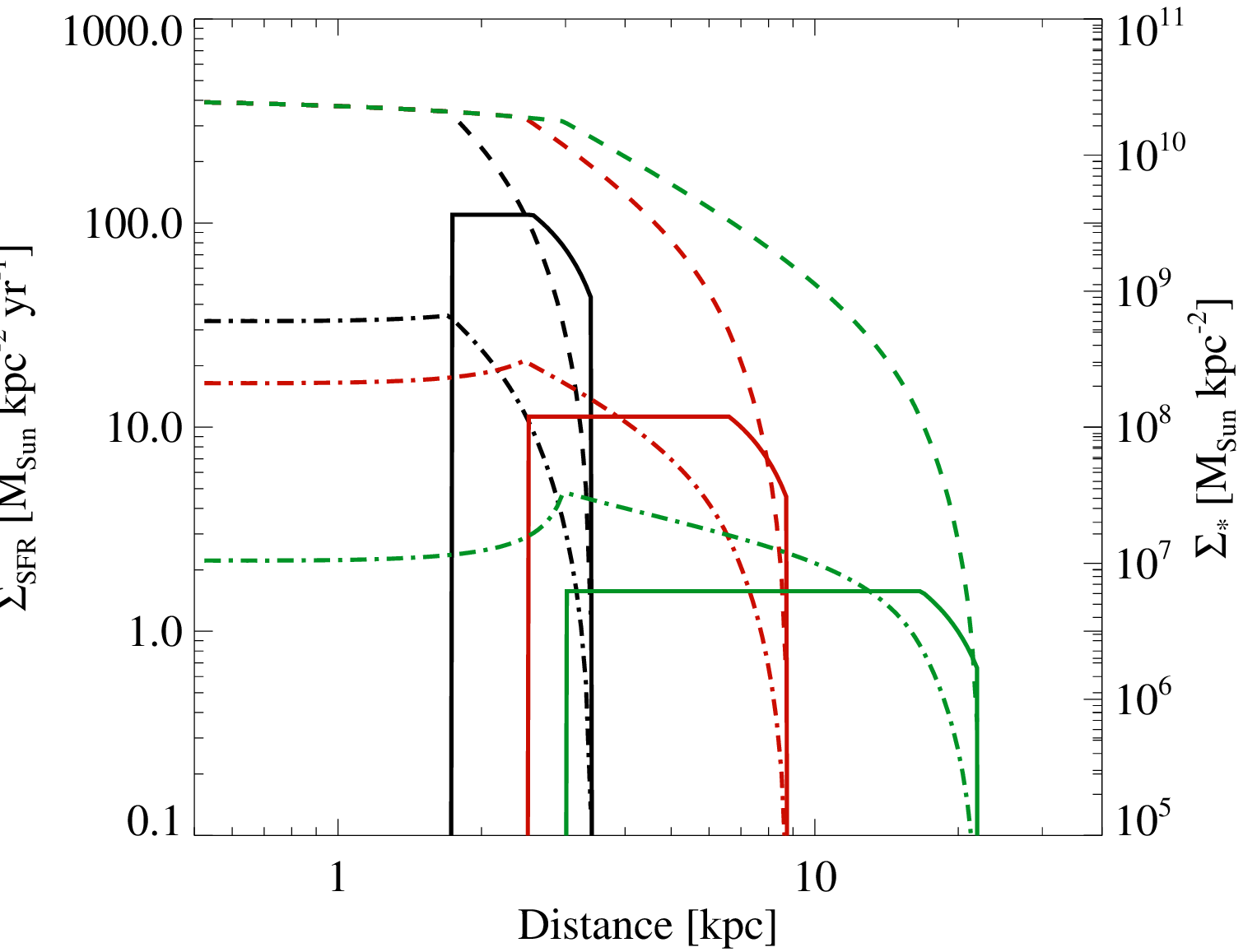}}
    \subfloat{\includegraphics[width=0.32\textwidth]{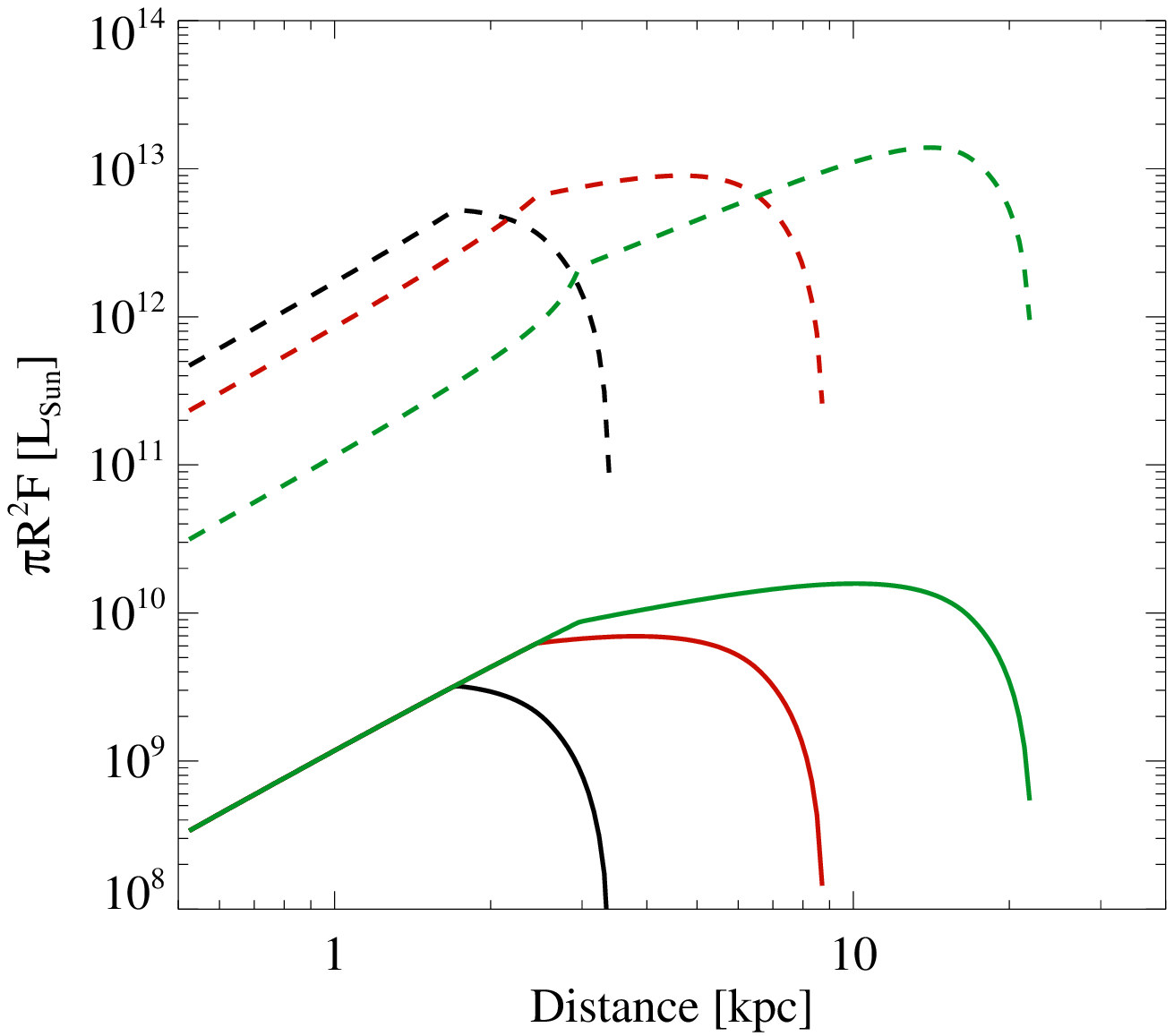}}
  \caption{Case 4, `Low-M0': Same as Figure \ref{fig:sfr1}, but
    $M_{\rm BH,0} = 0.1M_{\sigma}$. {\it Left panel}: Time evolution
    of star formation rate (green dashed line, scale on the left),
    total mass of gas converted into stars (black solid line) and
    luminosity of the newly formed stellar population (red dot-dashed
    line; both scales on the right). {\it Middle panel}: Radial plots
    at $3$, $10$ and $30$~Myr (black, red and green lines,
    respectively) of star formation rate density (solid, scale on the
    left), surface density of low-mass stars (dashed) and massive
    stars (dot-dashed; both scales on the right). {\it Right panel}:
    radial plot of $\pi R^2 F_*$ of the low-mass (solid) and massive
    (dashed line) stars; colour coding identical to the middle panel.}
  \label{fig:sfr4}
\end{figure*}

\begin{figure*}
  \centering
    \subfloat{\includegraphics[width=0.32\textwidth]{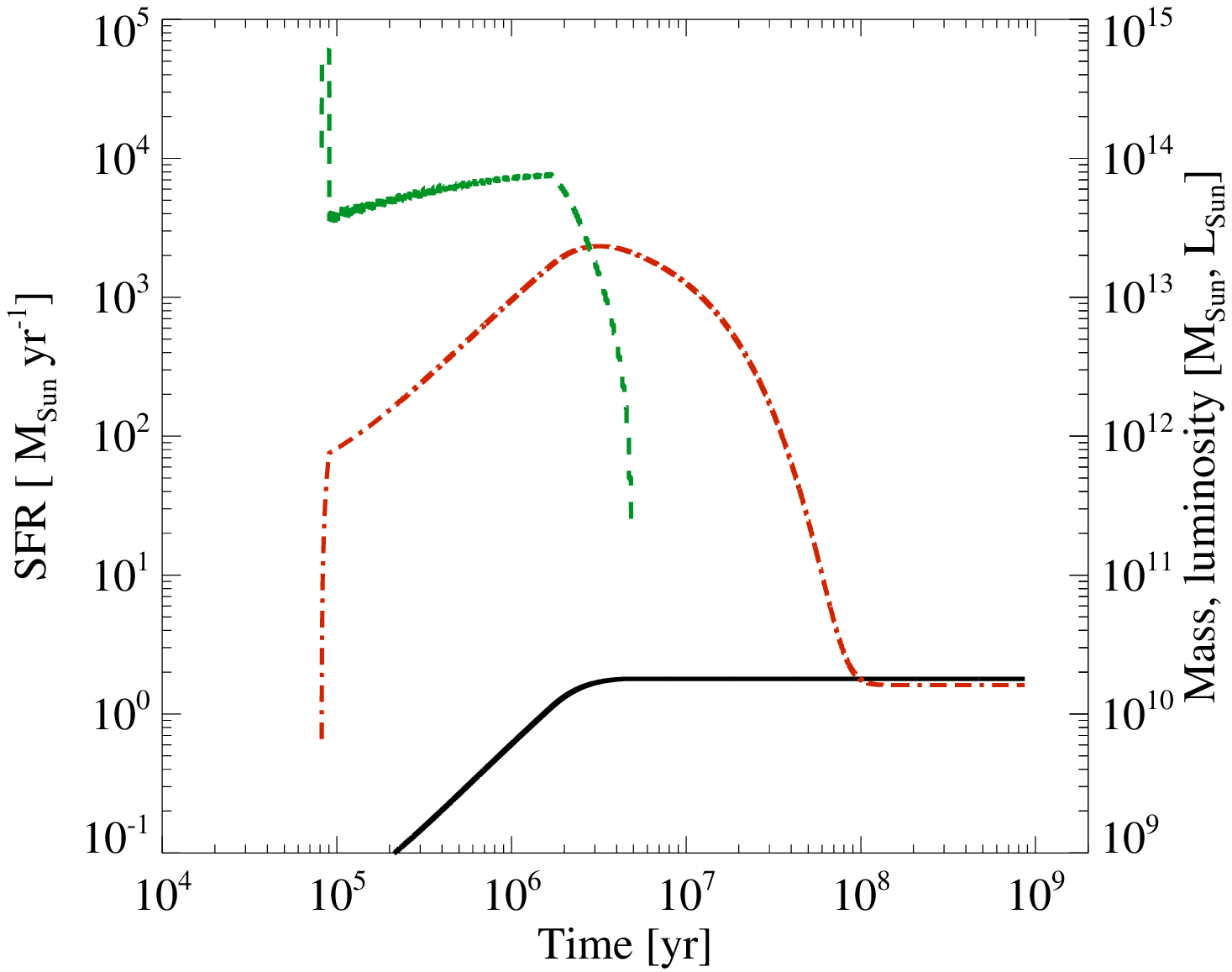}}
    \subfloat{\includegraphics[width=0.32\textwidth]{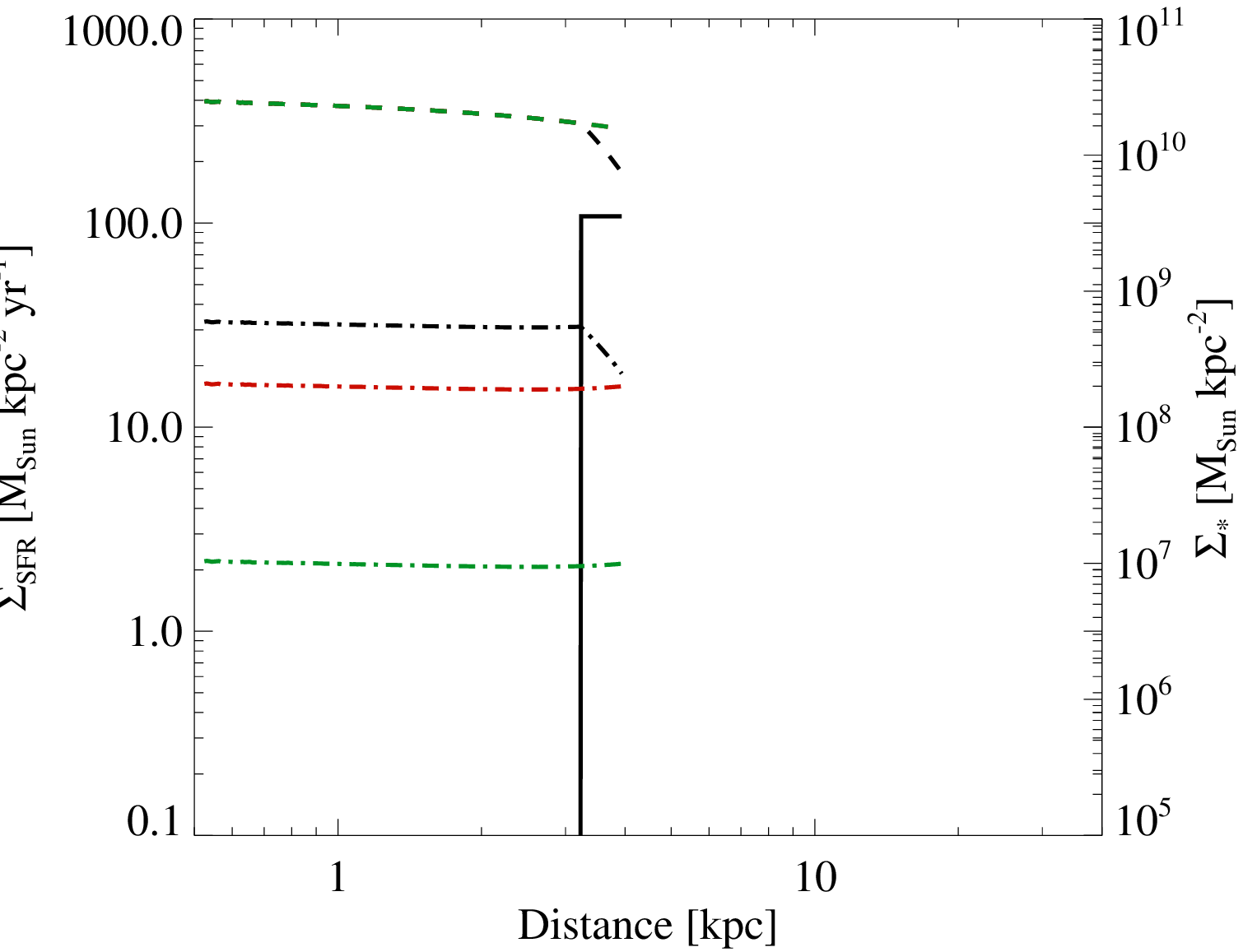}}
    \subfloat{\includegraphics[width=0.32\textwidth]{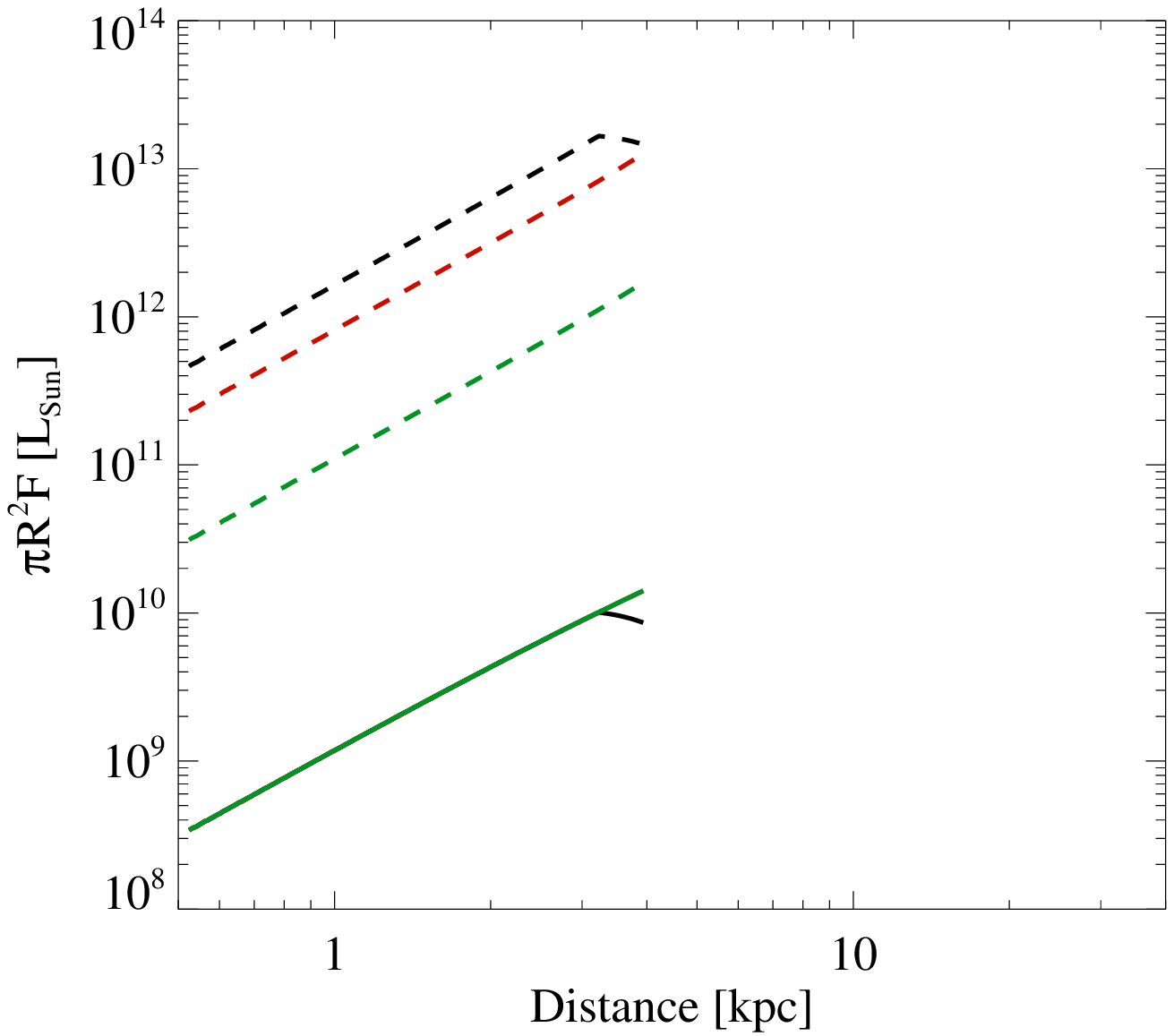}}
  \caption{Case 5, `Channel': Same as Figure \ref{fig:sfr1}, but
    outflow only affects the disc within $R < 4$~kpc. {\it Left
      panel}: Time evolution of star formation rate (green dashed
    line, scale on the left), total mass of gas converted into stars
    (black solid line) and luminosity of the newly formed stellar
    population (red dot-dashed line; both scales on the right). {\it
      Middle panel}: Radial plots at $3$, $10$ and $30$~Myr (black,
    red and green lines, respectively) of star formation rate density
    (solid, scale on the left), surface density of low-mass stars
    (dashed) and massive stars (dot-dashed; both scales on the
    right). {\it Right panel}: radial plot of $\pi R^2 F_*$ of the
    low-mass (solid) and massive (dashed line) stars; colour coding
    identical to the middle panel.}
  \label{fig:sfr5}
\end{figure*}

For each run, we present the results (Figures \ref{fig:sfr1} to
\ref{fig:sfr5}) in three plots. The first plot (left panel) shows the time
dependence of star formation rate $\dot{M}_*$ (green dashed line, scale on the
left; the jitter is caused by numerical effects), total mass of gas converted
into stars (black solid line, scale on the right) and the total luminosity of
the starburst (red dot-dashed line, scale on the right). The two other plots
show radial profiles of various quantities at 3, 10 and 30~Myr (black, red and
green lines, respectively). In the middle panels, the solid lines show the
star formation rate density (scale on the left), the dashed and dot-dashed
lines show the surface density of low-mass and massive stars, respectively. In
the right-hand-side panels, the quantity $\pi R^2 F_*$ is shown for low-mass
and massive stars (solid and dashed lines, respectively).

\subsection{Base run}

The physical processes occurring as the outflow washes past the disc are the
same as described analytically. Once the outflow moves past the cooling
radius, the disc is strongly compressed and begins forming stars; this happens
at around $10^5$~yr after the start of the AGN episode. The initial burst of
star formation is very strong, reaching almost $10^5 \; \msun$~yr$^{-1}$, due
to the high pressure just outside the cooling radius. Gas at this annulus is
rapidly consumed, however, and an approximately constant star formation rate
of $\sim 8 \times 10^3 \; \msun$~yr$^{-1}$ persists for $\sim 40$~Myr. During
this time, all of the disc gas ($M_{\rm g} \simeq 3 \times 10^{11} \; \msun$)
is converted into stars. The stellar luminosity, dominated by the massive
stars, increases with the converted mass, reaching a maximum of just over
$10^{14} \; \lsun$ at $\sim 30$~Myr and the drops as the massive stars die
out. After $10^8$~yr, the luminosity settles at $3 \times 10^{11} \; \lsun$
generated by the long-lived low-mass stars.

Although the total gas content is consumed in $40$~Myr, the consumption
timescale decreases for smaller radii. Looking at the middle panel of Figure
\ref{fig:sfr1}, we see that at any given time, star formation is occurring in a
ring, bounded on the inside by the radius within which all of the gas has
already been consumed by star formation (we call this the depletion radius
$r_{\rm dep}$) and on the outside by radius which the AGN outflow has
reached. The decreasing star formation density in the outer part of this ring
corresponds to the decreasing pressure in the shell of shocked ambient
material outside the contact discontinuity.

The density of low-mass stars, as well as their flux, follows the shape of the
original gas surface density in the disc within $r_{\rm dep}$ and decreases to
zero through the width of the star-forming ring. This, of course, does not
mean that there could not have been stars in the disc before the
outflow. Rather, the dashed lines represent the surface density of {\it young}
low-mass stars. The density of massive stars, on the other hand, decreases
with time as the stars die out. The radial profile of this density shows that,
especially at later times, a well-pronounced ring of young massive stars
develops around $r_{\rm dep}$, with few stars remaining well inside it and few
stars having been born well outside.

\subsection{Short AGN activity}

When the AGN switches off after $5$~Myr of activity (Figure \ref{fig:sfr2}),
its outflow begins to stall, eventually reaches some maximum radius and
subsequently collapses back on the galaxy. The pressure inside the outflow, as
well as the star formation rate, start decreasing immediately and fall by an
order of magnitude in $\sim 40$~Myr. The peak luminosity is reached earlier,
at $\sim 10$~Myr, and is a factor $\sim 2$ lower than in the `Base' run. Also,
some disc gas is never converted into stars - the total converted mass is only
$10^{11} \; \msun \simeq M_{\rm g} / 3$ at $10^8$~yr, a time beyond which our
calculation is no longer a fair representation of the physical processes
occurring as the outflow bubble collapses.

Radial plots of SF properties (middle and right panels) show that the
remaining disc gas is located on the outskirts. The central regions are still
efficiently converted into stars. The situation at $3$~Myr is, of course,
identical to the `Base' run, and even at $10$~Myr, there is very little
difference between the two cases. Only at $30$~Myr the difference becomes
significant, with both the inner and the outer edges of the star-forming ring
being much closer to the centre than in the `Base' run. The corresponding
stellar densities and fluxes are also lower.

\subsection{Lower star formation rate}

Reducing the star formation rate by a factor of $10$ (Figure \ref{fig:sfr3})
has the expected effect: the star formation rate and the total luminosity of
young stars drop by a similar factor. The peak of luminosity occurs at the
same time as in the `Base' run. However, the decline after the peak is slower,
owing to the larger amount of gas remaining throughout the disc. The sudden
drop of both these quantities at $4 \times 10^8$~yr correspond to the point
where the outflow velocity formally reaches $v_{\rm w} = 0.1 c$ and becomes
fixed at that value, leading to a drop in pressure.

The total mass of gas converted into stars is $\sim 0.25$ of the initial disc
gas mass. Gas is consumed very slowly throughout the disc; the middle panel
shows that between $3$ and $30$~Myr, only a ring $\sim 50$~pc wide is fully
consumed. Even at $10^8$~yr after the start of AGN activity, gas is completely
converted into stars only within $r < 800$~pc, with the fractional conversion
as low as $\sim 10\%$ at $r \simeq 10$~kpc. The stellar density and luminosity
plots follow the same trend and have significantly lower values than in the
`Base' run.

\subsection{Low initial SMBH mass}

When the black hole mass is 10 times lower than the $M_\sigma$ value,
i.e. $M_0 = 3.7 \times 10^7 \; \msun$, the evolution of the star formation
rate is surprisingly similar to the `Base' run (Figure \ref{fig:sfr4}). The
star formation rate is approximately proportional to the outflow pressure
which is, in turn, approximately proportional to $v_{\rm out}^2 \propto M_{\rm
  BH}^{2/3}$. Hence the star formation rate and the associated luminosity of
young stars are both approximately 5 times lower than in the `Base' run. The
radial profile plots reveal that the star formation rate at a given time is
roughly the same in both cases, but the star-forming ring is approximately
twice closer to the SMBH than in the `Base' run; this is again expected, as
$r_{\rm out} \propto v_{\rm out} \propto M_{\rm BH}^{2/3}$. The width of each
ring is almost identical in either case in logarithmic space. The fraction of
gas converted into stars is almost $100\%$, however, because the starburst
lasts longer.

\subsection{Channelled outflow}

Finally, we confine the effects of the AGN outflow to a region $r < 4$~kpc of
the disc. As might be expected, the evolution of the starburst follows the
`Base' run precisely until $\sim2$~Myr, when the outer shock of the outflow
reaches the confinement radius. Then the star formation rate declines very
rapidly, as gas in the central parts of the disc is consumed within
$5$~Myr. The stellar luminosity decrease is slightly slower, but by $10^8$~yr,
all the massive stars have died out and only the low-mass stars remain (see
the red dot-dashed curve in the left panel of Figure \ref{fig:sfr5}).

\subsection{Result summary}

Our semi-analytical model shows that the induced star formation in a galaxy
disc can be surprisingly varied, depending on just a few initial
parameters. Nevertheless, a common trend emerges that the disc can be strongly
affected and the starburst can consume a large fraction of the gas. This
implies that even though such powerful galactic outflows are rare in spiral
galaxies, they can have a lasting impact on the stellar population of the host
galaxy. We discuss the implications of these results in the next section.

\section{Discussion} \label{sec:discuss}

The current paradigm of AGN connection to their host galaxies is based on {\it
  negative} feedback, i.e. AGN activity is considered as the agent quenching
star formation in its host, by expelling the gas on short timescales. Our
results indicate that this is not the full picture. While it is true that an
energy-driven (i.e. non-cooling) AGN outflow clears diffuse gas out of the
bulge and halo of its host galaxy, the effect on the galaxy disc and any dense
structures is opposite. Soon after the AGN switches on and begins driving a
large-scale outflow, the pressure in the outflowing gas initiates a starburst in
the disc (see discussion in Section \ref{sec:compress}). Therefore we expect
that bright AGN in spiral galaxies, capable of driving energy-driven outflows
out to large distances in their hosts, are accompanied by starbursts in the
galaxy discs. We predict a sub-linear scaling between starburst and AGN
luminosities (Section \ref{sec:lumdiscuss}) and a number of morphological
properties of the starburst (Section \ref{sec:morphology}) that are consistent
with current observations (Section \ref{sec:evidence}). We review these points
in detail below.

In general, one may expect a correlation between AGN and
  starbursts simply because both are fed by gas within the galaxy and
  so more gas-rich galaxies are more likely to have both starbursts
  and AGN. However, such a correlation cannot be very tight, because
  AGN must be fed by small angular momentum gas \citep{Hobbs2011MNRAS},
  whereas star formation can easily take place in gas with very large
  angular momentum, e.g., in the galactic disc with length scale $>
  1$~kpc. More specifically, \citet{Nayakshin2012ApJ} argued that AGN
  cannot be fed very efficiently via galaxy discs in which case AGN
  accretion and star formation in the disc are completely
  decoupled. They based their argument on the observations of
  pseudo-bulge galaxies that show under-weight blackholes
  \citep{Hu08,Graham08} which poorly correlate with their host's
  properties \citep{KormendyEtal11}. \citet{Nayakshin2012ApJ} suggest
  that galaxy discs in the systems with under-weight SMBH are turned
  into stars much more rapidly than their SMBH could grow. This
  exemplifies how AGN activity and star formation could be decoupled.

We comment on the importance of various starburst properties and
  how they can help distinguish between `coincidental' and triggered
  starbursts below.

\subsection{Observational evidence of triggered star formation} \label{sec:compress}

A major component of our model is the connection between external
pressure in an AGN outflow and enhanced star formation in the galaxy
disc.  Observations of many star-forming regions show that the
fraction of gas converted into stars per dynamical time is a few
percent in `normal' (i.e. average density) environments, but rises in
regions of galaxy discs where the gas surface density $\Sigma \simgt
100 \; \msun$~pc$^{-2}$ \citep{Bigiel2008AJ}. There have been two
  conflicting interpretations of this phenomenon. One argument is that
  at this density, diffuse gas pressure upon molecular clouds starts
  to dominate over the turbulent pressure of the clouds themselves,
  leading to cloud confinement and enhanced star formation
  \citep{Krumholz2009ApJ}. This model is based on the fact that
  external pressure affects two of the three main parameters governing
  the star formation rate: molecular hydrogen fraction and free-fall
  timescale in the clouds \citep{Krumholz2009ApJ}. Higher external
  pressure compresses the gas to higher density, which enhances
  cooling and thus formation of molecules, and shortens the gas
  free-fall time. Other models claim that simple self-regulation of
  star formation by the interplay of turbulence and self-gravity
  explains the increased SF efficiency without requiring external
  pressure \citep{Faucher2013arXiv}. Importantly, however, the latter
  model assumes that the diffuse ISM pressure upon clouds is
  negligible. In our model, the external pressure always dominates
  above the cloud turbulent pressure ($p_{\rm turb} \sim 10^{-11} -
  10^{-10}$~dyn cm$^{-2}$), so the model of \citet{Krumholz2009ApJ}
  may still be applicable.

On the other hand, extensive observations of star forming clouds in M82
  \citep{Keto2005ApJ} revealed that star formation happens only in
  clouds or even sections of clouds that are compressed by the diffuse
  ISM, while uncompressed clouds or sections of clouds do not show
  star formation. The inward motions of compressed clouds are highly
  supersonic and hence consistent with shock waves driven by external
  pressure but inconsistent with pure gravitational contraction. This
  dichotomy between compressed star-forming and uncompressed quiescent
  clouds is independent of cloud density either, further revealing
  that cloud self gravity is not the dominant trigger of star
  formation.

Molecular clouds have been observed that are both affected by
radiation pressure and ionization \citep{Sugitani1989ApJ,
  Sugitani1991ApJS, Sugitani1994ApJS} or shocks from nearby supernova
explosions \citep{Preibisch1999AJ} and have higher star formation
efficiency than their unaffected counterparts. Stars formed due to
such triggering also have more uniform ages \citep{Preibisch1999AJ}
than typical young star clusters.

Effects of nearby external sources, such as supernova shocks and/or radiation
from massive stars, on the star formation process have been investigated in
great detail \citep{Klein1980SSRv, Bertoldi1989ApJ, Kessel-Deynet2003MNRAS,
  Dale2007MNRASb, Bisbas2011ApJ}. It is now well known that both blast waves
from supernova explosions and ionizing radiation from massive stars can
compress gas in molecular clouds, enhancing its density and making collapse
more likely. Although some details of this radiation-driven implosion of
molecular clouds are different from our model, the central tenet that
increased external pressure leads to increased star formation rates is well
supported by observations.

\subsection{Observational appearance of an AGN-triggered starburst}

\subsubsection{Starburst morphology} \label{sec:morphology}

The morphology of the outflow and the resulting starburst depends on the
distribution of cold gas in the galaxy itself. In the simplest model of an
axially symmetric spiral galaxy, the outflow is also axially symmetric and
propagates radially outward from the galaxy's centre. Assuming that all the
cold gas in the galaxy resides in a disc, young stars are born in a disc as
well, with a ring of ongoing star formation surrounding its outskirts. As the
starburst proceeds, massive stars in the central regions begin to die out and
the luminosity decreases there. Eventually, the luminosity of the disc edges
becomes higher than in the middle, although this effect is weak and depends
strongly on the underlying cold gas density, so would be difficult to detect
(see the dot-dashed lines in the middle panels of all result plots).

Real galaxies, however, are not completely axisymmetric, and the finer
detail of galactic disc structure may be imprinted in the starburst
shape. Molecular gas is concentrated in large cloud complexes in
spiral arms \citep{Reuter1996A&A, Brouillet1998A&A}, so most of the
star formation happens in these clumps. Nevertheless, the AGN
  outflow propagates through the halo rather than the disc, and so the
  outflow properties are independent of disc morphology. As a result,
  the outflow moves with approximate axial symmetry, so the starburst
is still confined to clumps within a circular region centred on the
AGN.

Depending on relative densities and masses of the molecular cloud regions, gas
consumption (and, hence, star formation) timescales may vary strongly between
them, so that the details of starburst morphology become irregular. In this
case, the regions of ongoing star formation and clusters of young stars may be
distributed asymmetrically throughout the starburst region.

\subsubsection{Star formation efficiency} \label{sec:sfe}

In our models, the fraction of gas converted into stars is typically very
large, $\sim 1$. In the `Base' model, all of the gas is converted in a few
times $10^7$~yr, less than a dynamical time in the galaxy ($t_{\rm d} \sim
10^8$~yr). `Low-M0' converts all the gas in approximately one dynamical
time. We see that AGN-triggered star formation can be very efficient.

There are several possible complications to this simple picture. First of all,
AGN activity episodes only last a few to $\sim 100$~Myr
\citep{Rawlings1991Natur}. We consider the effect of a short activity episode
in the `Low-tq' model. There, we find that $\sim 50\%$ of the gas is converted
into stars after one dynamical time. Thus even a short burst of AGN activity
can significantly affect the disc of its host galaxy.

A further complication arises due to the small-scale effects, such as
turbulence, magnetic fields, non-uniform feedback from massive stars and the
multiphase nature of the disc gas, affecting the star formation rate in any
given region. We estimate their importance in the model `Low-SFR', where star
formation rate is reduced by a factor 10. Even there, $\sim 16\%$ of the gas
is converted into stars after one dynamical time, much more than the few per
cent typical of ``standard'' star formation regions.

While the uncertainties involved in the calculation of the actual star
formation efficiency prevent us from making definitive quantitative statements
regarding the star formation efficiency, we believe that a significant
increase in SFE above that of undisturbed star formation regions is a robust
conclusion.

\subsubsection{Starburst luminosity} \label{sec:lumdiscuss}

The ratio of starburst and AGN luminosities determines, to a large extent, the
appearance of the galaxy in different wavebands. The analytically predicted
ratio (eq. \ref{eq:lratio}), $L_* \sim 4.5 L_{\rm AGN}$ and the results of
numerical calculations ($L_* = 10^{12}-10^{14} \; \lsun$, to be compared with
the Eddington luminosity $L_{\rm Edd} \sim 2 \times 10^{13} \; \lsun$ for an
SMBH on the $M-\sigma$ relation) agree well with with the observed AGN
contribution to total ULIRG luminosity, which is $\sim 20\%$
\citep{Genzel1998ApJ}.

The sub-linear scaling of starburst luminosity with AGN luminosity ($L_*
\propto L_{\rm AGN}^{5/6}$, see Section \ref{sec:lum}) can be understood in
terms of AGN outflow propagation. Lower mass (and lower luminosity) SMBHs take
longer to develop host-sweeping outflows since the outflow energy production
rate is proportional to $L$. However, outflows then linger in the host for
longer, since their velocity is proportional to $L^{1/3}$. Therefore small or
low luminosity AGN remain relatively more effective in triggering starbursts
than could be naively expected.

\subsubsection{Stellar ages}

An energy-driven AGN outflow moves with a velocity $v_{\rm e} \sim 1000$~km/s
and passes through the galaxy disc in several tens of Myr. Significant amounts
of gas are consumed within the affected parts of the disc on a similar
timescale. Even if the AGN switches off, the outflow persists for more than an
order of magnitude longer than the driving phase \citep[see also Figure
  \ref{fig:sfr2}]{King2011MNRAS} and gas is consumed efficiently. Therefore
the ages of stars formed during the starburst differ by $\simlt 100$~Myr. Such
a uniform-age stellar population throughout the galaxy is a potentially strong
indicator of an AGN-induced starburst in the past. Recent observations of the
star cluster population in M82 \citep{Lim2013arXiv} reveal a large number of
clusters with a mean age of $500$~Myr found everywhere in the galaxy disc. Our
model predicts that these cluster were created during and just after an AGN
activity episode $\sim 500-600$~Myr ago. M82 is known to have interacted
tidally with a companion galaxy M81 $\sim 600$~Myr ago \citep{deGrijs2001AJ};
this interaction may have caused a rapid gas inflow into the central regions
of M82, triggering an AGN episode.

This simple picture can be complicated by several effects. First of all, AGN
activity episodes may recur every $\sim 100$~Myr and produce a new
starburst. These starbursts should be weaker, however, since the gas density
in the spheroidal components of the galaxy decreases significantly after the
first outflow clears it, therefore subsequent outflows have lower
pressure. Nevertheless, the star formation history of a galaxy would show
several peaks with spreads that may be comparable to the duration between
them, blurring the stellar age profiles. In addition to this, subsequent
activity episodes may be triggered by the same gas that was once removed from
the galaxy cooling and recollapsing; in that case, the outflow has a similar
gas density to the previous one, further obscuring the bursty star formation
history of the galaxy.

\subsection{Comparison with merger-triggered starbursts}

It is well-known that galaxy mergers can also fuel starbursts by
mixing gas, creating turbulence and shocks and altering the
gravitational potential of the merging system. These processes create
a complex and irregular morphology of the galaxy and the starburst,
without any easily identified structure. Careful analysis of
  morphological differences should help distinguish between
  merger-induced and AGN-triggered starbursts (see Section
\ref{sec:morphology}).

First of all, the morphology of the whole galaxy may indicate the
  origin of the starburst. In our model, the starburst happens in a
  galaxy disc or dense structures engulfed by the outflow. Major
  galaxy mergers usually leave elliptical remnants. However,
  \citet{Springel2005ApJ} showed that mergers of extremely gas-rich
  spirals preserve galactic discs. Therefore a galaxy with a disc
  undergoing a starburst was not necessarily induced by an AGN. Minor
  mergers do not destroy galaxy discs either, but cause compact
  starbursts and/or fuel the AGN. These starbursts, however, are
  likely to be asymmetric due to the asymmetric interaction between
  the galaxy and its satellite, and therefore they should be
  distinguishable for AGN-triggered starbursts.

If the minor merger triggers AGN activity, the resulting outflow passes
through disturbed material and the AGN-triggered starburst can be as irregular
in the central parts of the galaxy as the one triggered directly by the
merger. The tell-tale sign of AGN-induced star formation activity in this case
may be outward velocities of young stars in regions directly exposed to AGN
feedback, as shown in \citet{Nayakshin2012MNRASb} and
\citet{Zubovas2013MNRAS}. In those papers, we studied the impact of the AGN
outflow ram pressure on the outflowing gas, finding that the outer outflowing
shell itself can fragment and produce stars. In an irregularly shaped galaxy,
the outflowing gas encounters many dense clumps, which can be accelerated and
even ejected from the galaxy. In principle, isotropic ejection of stars, as
opposed to tidal tails from merging galaxies, could be used to distinguish
between the two starburst causes, but in practice this would require very
detailed observations and identification of stellar orbits, which is not
currently feasible.

Finally, in the cases where a galaxy merger triggers both a starburst and an
AGN, the latter usually follows the former with a delay of $\sim 10^8$~yr
\citep{Schawinski2007MNRAS}. In our model, the opposite is true - the
starburst begins a few Myr after the AGN turns on. While it is difficult to
measure the ages of AGN, in principle this difference could be used to
distinguish between the two regimes. A recent discovery of a high-redshift QSO
with a nearby cluster of young stars \citep{Rauch2013arXiv} shows that
AGN triggering of star formation may be detected.

\subsection{Observational evidence for AGN-triggered star formation} \label{sec:evidence}

Direct observational evidence of starbursts triggered by AGN activity is
difficult to find due to briefness of AGN phases, uncertainty of the age of
any given AGN and the confusion between AGN and starburst contributions to
total galaxy luminosity.

Many Ultra-luminous IR galaxies (ULIRGs) and Narrow-line Seyferts (NLSs) show
both starburst and AGN activity
\citep{Genzel1998ApJ,Farrah2003MNRAS,Sani2010MNRAS} and starbursts typically
have ages between $10^7$ and $10^8$~yr in the nuclear regions
\citep{Genzel1998ApJ}, which is consistent with typical AGN activity durations
\citep{Rawlings1991Natur,Martini2001ApJ,Hopkins2005ApJ}. However, this only
shows that AGN and starbursts are coincident, but does not reveal which one
causes, or even precedes, the other \citep{Lutz1998ApJ}.

Some galaxies hosting both starbursts and AGN are at least consistent with our
model. For example, the starburst luminosity in Seyfert galaxies increases
with SMBH mass and luminosity \citep{Sani2010MNRAS}; we predict the same (see
Section \ref{sec:lumdiscuss}). In fact, our predicted relation $L_* \propto
L_{\rm AGN}^{5/6}$ is very similar to observed correlations: $L_* \propto
L_{\rm AGN}^{0.7}$ \citep{Sani2010MNRAS} and $L_* \propto L_{\rm AGN}^{0.8}$
\citep{Netzer2009MNRAS}.  The same observations also show starburst luminosity
increasing with increasing Eddington ratio of the AGN: $L_* / L_{\rm AGN}
\propto l$. Although our model (eq. \ref{eq:sflum}) predicts the opposite (if
weak) trend, this happens because $L_{\rm AGN}$, $l$ and the black hole mass
$M$ are not independent variables. When the relation is recast in terms of $M$
and $l$ \citep[as was done by][]{Sani2010MNRAS}, we find $L_* \propto
l^{2/3}$.

Recently, \citet{LaMassa2013arXiv} investigated a large sample of
active galaxies at various redshifts and found that star formation in
the central few kpc is positively, but sub-linearly, correlated with
AGN luminosity. This is consistent with our picture of a strong
starburst due to high outflow pressure close to the SMBH, decreasing
in strength as the outflow moves further out. On the other hand,
  central starbursts can easily be interpreted as fuelling the AGN
  rather than being triggered by the central source
  \citep{Thompson2005ApJ}.

Another piece of evidence comes from observations of radio-selected AGN that
also have starburst activity. \citet{Karouzos2013arXiv} show that in a sample
of more than a hundred radio AGN, the more luminous ones show more star
formation than would be expected from standard models based on galaxy
properties \citep{Elbaz2011A&A}. Also, galaxies with lower radio loudness
(i.e. more likely to be accreting in the `quasar-mode') also show higher star
formation rates than would be expected from empirical models of galaxy
evolution. Finally, in the same sample, AGN contribution to the total
bolometric luminosity of the sources is typically $40-60\%$, with a slight
increase toward brighter AGN. All together, these properties are consistent
with our predictions in much the same way as observations of Seyfert galaxies
are.

Finally, there is some evidence that ULIRGs have more compact H II regions
than other types of star-forming galaxies \citep{Lutz1998ApJ}; this could
happen if gas in these galaxies is strongly compressed.

We conclude from this section that observational evidence of AGN triggering
star formation in spiral galaxies is inconclusive, but generally consistent
with the predictions of our model.

\subsection{{\it Fermi} Bubbles -- an outflow in a gas-poor galaxy}

The model described in this paper is mainly applicable to outflows in gas-rich
spiral galaxies. Once such an outflow passes, however, the galaxy is left
gas-poor: bulge and halo gas has been expelled, while disc gas has been turned
into stars (see Section \ref{sec:sfe}, above). Our Milky Way is an example of
such a gas-poor spiral galaxy. An energy-driven AGN outflow may still occur in
these galaxies; it moves much faster, and in the Milky Way may have been the
cause of the {\it Fermi} bubbles \citep{Zubovas2011MNRAS,Zubovas2012MNRASa}.
This outflow did not cause a large-scale starburst in the galaxy disc, because
both the density in the outflowing material was much lower than envisioned
here, and the AGN luminosity powering it was low enough (due to the low mass
of \sgra) to make the pressure too low to cause a noticeable effect. The gas
fraction was $f_{\rm g} \sim 10^{-3}$ and the velocity was $v \sim 1000$~km/s,
creating a pressure 2 orders of magnitude lower than in the disc of a typical
gas-rich galaxy that we considered in this paper. This pressure becomes
comparable to the typical ISM pressure once the outflow gets out to a few kpc,
so any effect it may have is confined to the central regions of the
galaxy. Indeed, in our simulations of the {\it Fermi} bubbles
\citep{Zubovas2012MNRASa}, we find that the $200$-pc-wide Central Molecular
Zone of the Galaxy is perturbed, leading to gravitational fragmentation and
possible star formation. Other gas-poor galaxies would show similar effects
after AGN outflows; their presence may be used to infer past AGN activity even
when the more direct evidence is gone.

\subsection{Comparison to other work}

Positive effects of AGN feedback on the host galaxy's stellar content have
been investigated several times in the past few years. \citet{Silk2005MNRAS},
\citet[both analytical investigations]{Silk2009ApJ} and \citet[numerical
  simulations]{Gaibler2012MNRAS} consider a jet-driven outflow from an active
SMBH, which pressurises the ISM and increases its turbulence, leading to a
burst of star formation. Depending on the parameters of the ISM, the jet can
either produce a starburst along a narrow beam or a more wide-spread one,
encompassing most of the galaxy. Numerical simulations give jet-induced star
formation rates of a few hundred $\msun$~yr$^{-1}$ distributed in an expanding
ring and limited by diffuse gas removal from the galaxy by the
jet. \citet{Silk2005MNRAS} does not provide estimates for star formation rate,
but considers a spherical geometry, so that model is appropriate for
starbursts in elliptical galaxies.

\citet{Ciotti2007ApJ} use high-resolution numerical simulations to investigate
AGN feedback effect on ISM thermodynamics, with effects on star formation
being a secondary result. They find that radiative AGN feedback can trigger
repeated nuclear starbursts, again with star formation rates of several
hundred $\msun$~yr$^{-1}$, but not necessarily related to large-scale
outflows. The starbursts in our model are larger in both physical extent and
SFR, so a direct comparison is not warranted.

\citet{Ishibashi2013arXiv} consider a model where a momentum-driven AGN
outflow shell expands and simultaneously forms stars at some efficiency. They
use this prescription to explain the inside-out growth of galaxies since $z
\sim 2$, as these shells deposit stars preferentially at large radii. However,
since the model is very simplistic and does not deal with dense structures
compressed by the outflow, we do not compare our results to theirs directly.

\section{Summary and conclusion} \label{sec:concl}

We have shown, using analytical arguments and a semi-analytical model, that
powerful AGN outflows in spiral galaxies may have a strong positive effect on
the galaxy disc. The disc gas is compressed by the outflow as the latter
passes on either side of the disc. This leads to an enhancement in star
formation rates and causes most of the gas in the disc to be turned into
stars. The starburst continues for several to several tens of Myr even after
the AGN switches off, provided there is enough gas left in the disc.

Observationally, such a galaxy would appear as a ULIRG with an almost flat
central young star luminosity density, with this disc surrounded by a ring of
ongoing star formation. The AGN component might exist, but would be
subdominant to the total luminosity of the starburst. These properties are
consistent with some observations of ULIRGs, however it is very difficult to
disentangle the causal connection between a starburst and an AGN and as such,
it may be some time before strong evidence in support of this model is found.

It is interesting to consider our results in the broader context
  of the effect of AGN activity on star formation in the host
  galaxy. The difference between positive and negative feedback is
  rather clear in the spheroids of quiescent galaxies: gas-rich
  outflows cool and fragment, forming stars
  \citep{Nayakshin2012MNRASb}, while gas-poor outflows simply clear
  the remaining gas from the galaxy \citep{Zubovas2012ApJ}. This
  picture gets somewhat more complicated in mergers, where the outflow
  would propagate in a disturbed medium, compressing gas and
  trigerring fragmentation in some directions while blowing gas away
  in others. The current results show that whichever way feedback
  affects gas in the galaxy bulge and halo, the disc gets compressed
  and undergoes a starburst. The star formation rate, $\dot{M}_*
  \lesssim 5 \times 10^3 \; \msun$~yr$^{-1}$, is comparable to the
  mass outflow rate in the spherical energy-driven outflow,
  $\dot{M}_{\rm out} \simeq 2 \times 10^3 \; \msun$~yr$^{-1}$
  \citep{Zubovas2012ApJ}. Therefore it appears that the net effect of
  an energy-driven AGN outflow is a quickening of galaxy evolution,
  rather than a strictly negative or strictly positive feedback.

\section*{Acknowledgments}

Theoretical astrophysics research in Leicester is supported by an STFC Rolling
Grant. KZ thanks the UK STFC for support successively in the form of a
studentship and a postdoctoral research associate position. MIW acknowledges
the Royal Society for financial support.

\bibliographystyle{mn2e}
\bibliography{../../zubovas}

\end{document}